\documentclass[aps,
prx,twocolumn,amsmath,amssymb,showpacs,superscriptaddress,notitlepage]{revtex4-1}
\usepackage{amsmath,amssymb,amsfonts,bm}
\usepackage{graphicx}
\usepackage{float}
\usepackage{subfigure}
\usepackage{epstopdf}
\usepackage{dcolumn}
\usepackage{mathrsfs}
\usepackage{bbold}
\usepackage{dsfont}
\usepackage[colorlinks=true,linkcolor=blue,citecolor=blue, urlcolor=blue,bookmarks=false]{hyperref}
\usepackage{changes}
\colorlet{Changes@Color}{red}

\begin{document}

\title{Layer Hall effect induced by hidden Berry curvature in antiferromagnetic insulators}
\author{Rui Chen}
\thanks{They contribute equally to this work. }
\affiliation{Shenzhen Institute for Quantum Science and Engineering and Department of Physics, Southern University of Science and Technology (SUSTech), Shenzhen 518055, China}
\affiliation{International Quantum Academy, Shenzhen 518048, China}

\author{Hai-Peng Sun}
\thanks{They contribute equally to this work. }
\affiliation{Shenzhen Institute for Quantum Science and Engineering and Department of Physics, Southern University of Science and Technology (SUSTech), Shenzhen 518055, China}
\affiliation{Institute for Theoretical Physics and Astrophysics, University of W\"urzburg, 97074 \"Wurzburg, Germany}

\author{Mingqiang Gu}
\thanks{They contribute equally to this work. }
\affiliation{Shenzhen Institute for Quantum Science and Engineering and Department of Physics, Southern University of Science and Technology (SUSTech), Shenzhen 518055, China}

\author{Chun-Bo Hua}
\affiliation{School of Electronic and Information Engineering, Hubei University of Science and Technology,
Xianning 437100, China}

\author{Qihang Liu}
\email{Corresponding author: liuqh@sustech.edu.cn}
\affiliation{Shenzhen Institute for Quantum Science and Engineering and Department of Physics, Southern University of Science and Technology (SUSTech), Shenzhen 518055, China}
\affiliation{Guangdong Provincial Key Laboratory of Computational Science
and Material Design, Southern University of Science and
Technology, Shenzhen 518055, China}

\author{Hai-Zhou Lu}
\email{Corresponding author: luhz@sustech.edu.cn}
\affiliation{Shenzhen Institute for Quantum Science and Engineering and Department of Physics, Southern University of Science and Technology (SUSTech), Shenzhen 518055, China}
\affiliation{International Quantum Academy, Shenzhen 518048, China}
\affiliation{Shenzhen Key Laboratory of Quantum Science and Engineering, Shenzhen 518055, China}

\author{X. C. Xie}
\affiliation{International Center for Quantum Materials, School of Physics,
Peking University, Beijing 100871, China}
\affiliation{Collaborative Innovation Center of Quantum Matter, Beijing 100871, China}
\affiliation{CAS Center for Excellence in Topological Quantum Computation,
University of Chinese Academy of Sciences, Beijing 100190, China}

\begin{abstract}
The layer Hall effect describes electrons spontaneously deflected to opposite sides at different layers, which has been experimentally reported in the MnBi$_2$Te$_4$ thinfilms under perpendicular electric fields [Gao \emph{et al.}, \textcolor{blue}{Nature \textbf{595}, 521 (2021)}].
Here, we reveal a universal origin of the layer Hall effect in terms of the so-called hidden Berry curvature, as well as material design principles. Hence, it gives rise to zero Berry curvature in momentum space but nonzero layer-locked hidden Berry curvature in real space. We show that compared to that of a trivial insulator,  the layer Hall effect is significantly enhanced in antiferromagnetic topological insulators.  Our universal picture provides a paradigm for revealing the hidden physics as a result of the interplay between the global and local symmetries, and can be generalized in various scenarios.
\end{abstract}
\maketitle

\section{Introduction}
\begin{figure}[h!]
\centering
\includegraphics[width=0.95\linewidth]{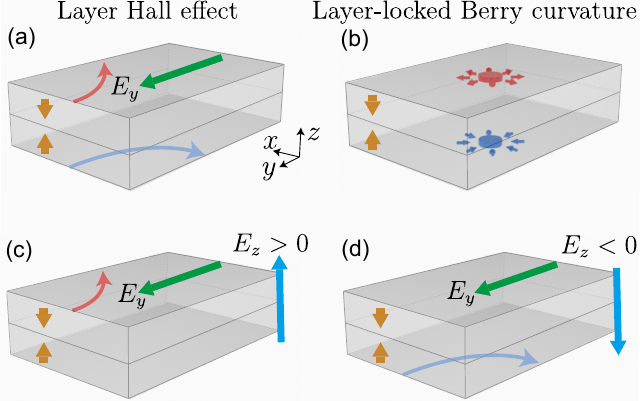}
\caption{Schematics of (a) the layer Hall effect and (b) layer-locked hidden Berry curvature (the red and
blue fluxes) in a two-layer antiferromagnetic (AFM) insulator. In the layer Hall effect, electrons are spontaneously deflected to opposite sides at different layers (the red and blue arrowed curves) due to the layer-locked hidden Berry curvature. (c-d) When a perpendicular electric field (the cyan arrow) is applied, the system breaks the $\mathcal{PT}$ symmetry ($\mathcal{P}$ for inversion and $\mathcal{T}$ for time-reversal), and shows layer-locked anomalous Hall effects tunable by the electric-field direction. The yellow arrows specify the AFM configurations. The green arrows denote the in-plane electric field $E_y$ for the Hall measurement.}
\label{fig_illustration}
\end{figure}

An electron has multiple degrees of freedom, including
charge, spin, and valley. Such degrees of freedom are encoded with distinct Berry curvature distribution, leading to various types of Hall effect including the anomalous~\cite{Nagaosa10rmp}, spin~\cite{Sinova15rmp,Bernevig06sci}, and valley \cite{Xiao07prl} Hall effects. Recently, the direct observation and manipulation of these Hall effects have already been achieved, triggering further explorations of the family of Hall effects~\cite{Kato04sci,Konig07sci,Roth09sci,Mak14sci,Chang13sci}. In addition to charge, spin, and valley, electrons possess another degree of freedom that divides real space, especially in layered materials, i.e., the layer degree of freedom. Accordingly, this may lead to an unprecedented type of Hall effect, the layer Hall effect, where electrons are spontaneously deflected to opposite sides at different layers [Fig.~\ref{fig_illustration}(a)] \cite{Gao2021Nature,ChenCZarXiv2022}. Such effect is required to be associated with a nonzero layer-dependent Berry curvature locked with real space [Fig.~\ref{fig_illustration}(b)], in contrast to the valley Hall effect in which the Berry curvature is inhomogeneously distributed in momentum space.

In this work, by using tight-binding model Hamiltonians as well as first-principles calculations, we study the layer Hall effect in an A-type antiferromagnetic (AFM) system that preserves the global space-time $\mathcal{PT}$ symmetry ($\mathcal{P}$ stands for inversion symmetry and $\mathcal{T}$ stands for time-reversal symmetry). We show that whereas the global Hall conductance vanishes because of the antiunitary $\mathcal{PT}$ symmetry, each layer that breaks the $\mathcal{PT}$ symmetry locally contributes to the nonvanishing layer-dependent Hall conductance, as a result of the layer-locked hidden Berry curvature. Remarkably, the layer Hall effect can be significantly enhanced in the ultrathin films of the AFM topological insulator, compared to that in a topologically trivial insulator. The layer Hall effect manifests a switchable net anomalous Hall conductance by applying a perpendicular electric field, which lifts the two-fold degeneracy of the states with the compensated layer-locked Berry curvature [Figs.~\ref{fig_illustration}(c-d)]. On the other hand, when the thickness increases, the enhanced layer Hall effect in the AFM topological insulator is localized to the top and bottom surfaces, with the Hall conductance approaching $\pm e^2/2h$. Considering that MnBi$_2$Te$_4$~\cite{Otrokov19nat,Deng20sci} is recently believed to host the axion-insulator phase but challenging to be detected~\cite{ChenCZ2019PRL,ChenR2020arXiv}, our theory is helpful for detecting the axion-insulator phase through nonlocal and standard Hall-bar measurements.

\section{Hidden Berry curvature and layer Hall effect in $\mathcal{PT}$-symmetric antiferromagnets}
\label{Sec_Berry}
It is recently recognized that various physical effects are determined by the local symmetry breaking of a system, albeit with a higher global symmetry that seemingly prohibits the effect from happening. Examples include spin polarization~\cite{Zhang2014NatPhys}, orbital polarization~\cite{Beaulieu2020PRL}, circular polarization~\cite{Razzoli2017PRL}, and Berry curvature~\cite{Cho2018PRL}, etc.
As a result, the concept of {\it hidden polarization} could be defined, where the specific physical quantity is localized in real space due to the local symmetry breaking, whereas the global symmetry ensures an energy-degenerate partner with opposite polarization.

To illustrate this, we consider a $\mathcal{PT}$-symmetric unit cell consisting of two sectors (say, layers) with broken $\mathcal{PT}$ for each. Examples include nonmagnetic 2H-stacking MoS$_2$ and A-type AFM insulator MnBi$_2$Te$_4$, no matter for 3D bulk or 2D few-layer slabs, where a single MoS$_2$/MnBi$_2$Te$_4$ unit breaks the $\mathcal{PT}$ symmetry locally.
The antiunitary nature of the global $\mathcal{PT}$ symmetry ensures that each band is two-fold degenerate with zero net Berry curvature
in momentum space $\Omega_{n,\uparrow}(k)+\Omega_{n,\downarrow}(k)=0$, where $n$ and $k$ denote the band index and wave vector, respectively.
However, each $\mathcal{PT}$-broken layer manifests a nonzero distribution of local Berry curvature,
$\Omega_n(k,z)$ ($z=1$ or 2 is the layer index), also named as hidden Berry curvature.
Meanwhile, another layer related by the $\mathcal{PT}$ symmetry manifests an opposite hidden Berry curvature distribution, i.e., $\Omega_n(k,1)=-\Omega_n(k,2)$.
For nonmagnetic material such as MoS$_2$, the integral of the hidden Berry curvature throughout the whole Brillouin zone (BZ) must be zero because the preserved $\mathcal{T}$ ensures $\Omega_n(k,z)=-\Omega_n(-k,z)$. As previously reported~\cite{Cho2018PRL,Schuler2020SciAdv}, such hidden physics can only be experimentally detected by special measurements with momentum resolution, such as circular-dichroism angle-resolved photoemission. However, in a $\mathcal{PT}$-symmetric A-type AFM insulator with broken $\mathcal{T}$, the integral of the hidden Berry curvature could give rise to a nonzero $\Omega_n(z)$, leading to measurable signals from standard measurements such as quantum oscillation or transports~\cite{Murakawa2013}.
Therefore, an even-layer A-type AFM insulator is an ideal platform to study the hidden Berry curvature and the layer Hall effect [see Figs.~\ref{fig_illustration}(a-b)].

\begin{figure*}[hptb]
\centering
\includegraphics[width=0.85\linewidth]{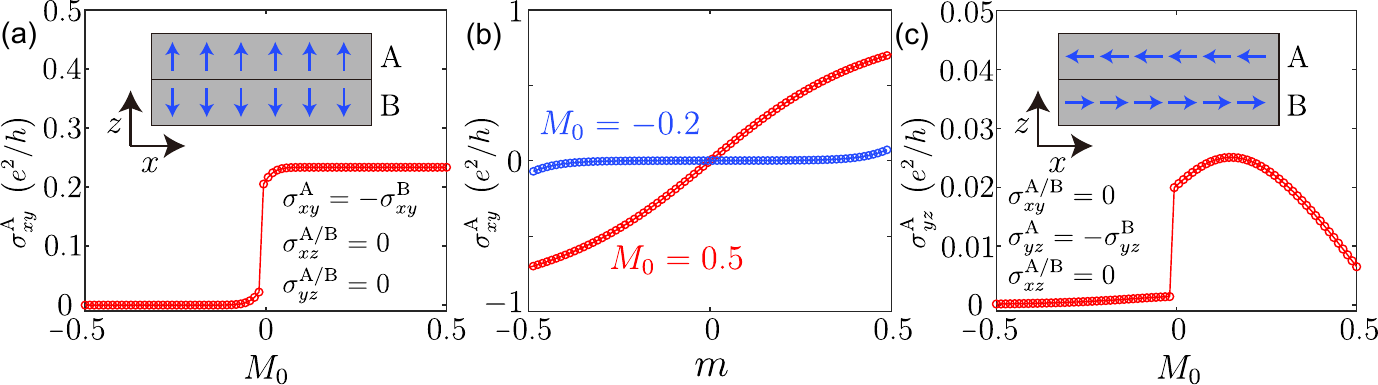}
\caption{Numerically calculated Hall conductance $\sigma_{xy}^{\text{A}}$ (a) as a function of $M_0$ with $m=0.12$ and (b) as a function of $m$ with $M_0=0.5$ (red) and $M_0=-0.2$ (blue). (c) Numerically calculated Hall conductance $\sigma_{yz}^{\text{A}}$ as a function of $M_0$ with $m=0.12$. Here we consider a 3D bulk AFM insulator. The system size is adopted as follows: (a-b) $n_x=n_y=40$ and $n_z=2$ and (c) $n_y=n_z=40$ and $n_x=2$ with periodic boundary conditions along the $x$, $y$, and $z$ directions.
}
\label{fig_conductance_bulk}
\end{figure*}

\section{Model and method}

We next choose an A-type AFM insulator to study the layer Hall effect and its dependence on the magnetization, band topology, and experimental signatures. We start from a tight-binding model with out-of-plane magnetization defined on a cubic lattice as~\cite{Zhang19prl,Zhang20prl,LiHailong21PRL,ChenCZ2019PRL,Ding20prbrc,ChenR2020arXiv,ChenCZ2021arXiv}
\begin{equation}\label{Hamiltonian}
H=\sum_{\mathbf{r}} \phi_{\mathbf{r}}^{\dagger} V_{\mathbf{r}} \phi_{\mathbf{r}}+\left(\sum_{\mathbf{r}, \alpha=x, y, z} \phi_{\mathbf{r}}^{\dagger} T_{\alpha} \phi_{\mathbf{r}+\delta \hat{\alpha}}+\text { H.c. }\right),
\end{equation}
with
$V_{\mathbf{r}}=\Gamma_4 \left(M_0- 2B_1-4B_2\right)+m \left(-1\right)^z  s_{z}\sigma_{0}+V\left[z-\left(n_z+1\right)/2\right]  s_{0}\sigma_{0}$, $T_x=-iA_2\Gamma_1/2+B_2\Gamma_4$, $T_y=-iA_2\Gamma_2/2+B_2\Gamma_4$, and $T_z=-iA_1\Gamma_3/2+B_1\Gamma_4$. Here, $\Gamma
_{j=1,2,3}=s_{i}\otimes \sigma _{1},$ $\Gamma _{4}=s_{0}\otimes \sigma _{3}$, and $\sigma$ and $\tau$ are Pauli matrices for the spin and orbital subspaces, respectively. $D_{i}$, $M_{0}$, $B_{i}$ and $A_{i}$ are model parameters, where $i=1,2$. $m$ describes the amplitude of the intra-layer ferromagnetic order, and $V$ measures
the potential on each layer induced by the perpendicular electric field $E_z$. In the absence of the perpendicular electric field, i.e., $V=0$, the system describes a trivial insulator when $M_0<0$, and an AFM topological insulator when $M_0>0$. In the calculations, we take the parameters as $A_1=A_2=0.55$ and $B_1=B_2=0.25$~\cite{LiHailong21PRL}.

The BZ-integrated hidden Berry curvature localized on each layer is obtained through the noncommutative real-space Kubo formula~\cite{Prodan2011JPA}
\begin{equation}
\Omega_n(z)= 2\pi i \operatorname{Tr} \left\{\hat{P}\left[-i\left[\hat{x}, \hat{P}\right],-i\left[\hat{y}, \hat{P}\right]\right]\right\}_{z},
\label{KuboFormula}
\end{equation}
with periodic boundary conditions along the $x$ and $y$ directions. Here, $\hat{x}$ and $\hat{y}$ are the coordinate operators and $\operatorname{Tr}\left\{\cdots \right\}_z$ is the trace over the $z$-th layer ($z=1,2,3,..., n_z$). $\hat{P}$ is the projector onto the occupied states of $H$.
Eq.~(\ref{KuboFormula}) is recognized as a local Chern marker representing the real-space projected Chern number in the $z$-direction \cite{PhysRevB.84.241106, Caio2019}. The layer-resolved Hall conductance is $\sigma_{xy}(z)=\frac{e^2}{h}\Omega_n(z)$, and the total Hall conductance is given by $\sigma_{xy}=\sum _z\sigma_{xy}\left(z\right)$.




\begin{figure*}[hptb]
\centering
\includegraphics[width=0.65\linewidth]{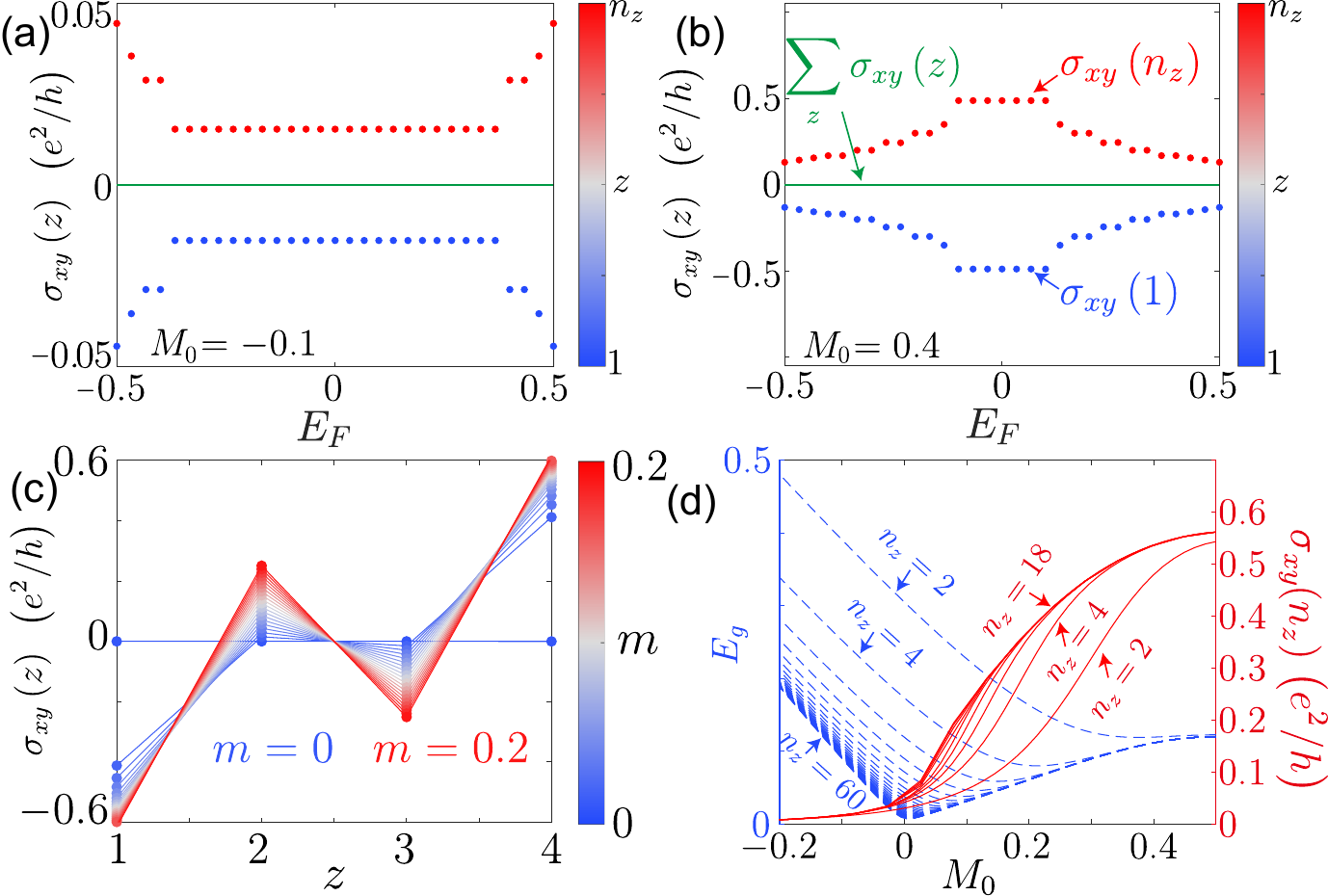}
\caption{(a-b) Numerically calculated Hall conductances as functions of the Fermi energy $E_F$ for (a) the trivial insulator film with $m=0.12$, and $M_0=-0.1$ and (b) AFM topological insulator thinfilm with $m=0.12$, and $M_0=0.4$ for $n_z=2$. Here the dots correspond to the layer-resolved Hall conductance $\sigma_{xy}\left(z\right)$ and the color distinguishes different layer $z$. The green lines denote the net Hall conductance $\sigma_{xy}=\sum _z\sigma_{xy}\left(z\right)$. (c) Numerically calculated Hall conductance $\sigma_{xy}\left(z\right)$ as functions of $z$ for different $m$ with $n_z=4$ and $M_0=0.4$. The color distinguishes different $m$. (d) Numerically calculated energy gap $E_g$ (dashed blue) and $\sigma_{xy}\left(n_z\right)$ (solid red) as functions of $M_0$ with $m=0.12$.  Here, we take $V=0$ with the system size $n_x=n_y=40$ and open boundary condition along the $z$-direction.
}
\label{fig_twolayer_V0}
\end{figure*}

\section{Layer Hall effect in the AFM insulator}
We first focus on the 3D A-type AFM insulator with the magnetic moment oriented along the $z$ axis [the inset in Fig.~\ref{fig_conductance_bulk}(a)]. With periodic boundary conditions along the $x$, $y$, and $z$ directions, we numerically calculated the layer-resolved Hall conductance $\sigma_{xy}^{\text{A/B}}$, which corresponds to the contribution to the net Hall conductance $\sigma_{xy}=\sigma_{xy}^{\text{A}}+\sigma_{xy}^{\text{B}}$ from the A/B layer. Despite that the net Hall conductance is zero for all the directions due to the $\mathcal{PT}$ symmetry, it is found that $\sigma_{xy}^{\text{A/B}}$ manifests nonvanishing layer-resolved Hall conductance as shown in Fig.~\ref{fig_conductance_bulk}(a). The global $\mathcal{PT}$ symmetry guarantees that the layer-resolved Hall conductances of the layers connected by the $\mathcal{PT}$ symmetry are exactly compensated, i.e., $\sigma_{xy}^{\text{A}}=-\sigma_{xy}^{\text{B}}$. Thus, the layers of the system can host opposite Hall conductances, which establishes the layer Hall effect.

When turned to an AFM topological insulator with a band inversion $(M_0>0)$, the layer Hall effect is significantly enhanced. As shown in Fig.~\ref{fig_conductance_bulk}(a), compared to the trivial insulator when $E_F$ is in the band gap, for the AFM topological insulator $\sigma_{xy}(n_z)$ reaches $0.23e^2/h$, which is more than one or two orders stronger.

The hidden Berry curvature as well as the layer Hall conductance experiences a steep increase during the topological phase transition, highlighting the significantly enhanced Berry curvature effects in topologically nontrivial systems. As shown in Fig.~\ref{fig_conductance_bulk}(b), with increasing magnetization $m$, $\sigma_{xy}^{\text{A}}$ increases both for the trivial insulators (blue) and the topological insulators (red), yet with different behaviors. The increasing rate of the AFM topological insulator ($M_0>0$) is much more prominent than that of the trivial insulator ($M_0 < 0$).

Figure~\ref{fig_conductance_bulk}(c) shows $\sigma_{yz}^{\text{A}}$ as a function of $M_0$ with the in-plane magnetic moment oriented along the $x$ axis [the inset in Fig.~\ref{fig_conductance_bulk}(c)]. Correspondingly, the nonzero layer Hall conductance turns to $\sigma_{yz}^{\text{A/B}}$ according to the manner of local $\mathcal{PT}$ breaking. Furthermore, the magnitude of the layer Hall effect with in-plane moments is much smaller compared to that with out-of-plane moments. This suggests that alignment between the magnetization and the stacking direction, which related to the partition of the unit cell into the $\mathcal{PT}$-breaking layers, is favorable to invoke a large layer Hall effect.

Let us now focus on the 2D case, i.e., bilayer system ($n_z=2$), composed of two layers with opposite magnetizations. Figure~\ref{fig_twolayer_V0}(a) shows the numerically calculated layer Hall conductance $\sigma_{xy}(z)$ as a function of the Fermi energy $E_F$ for a trivial insulator with $M_0=-0.1$. When turned to the bilayer of an AFM topological insulator with band inversion $(M_0=0.4)$, the layer Hall effect is significantly enhanced. As shown in Figs.~\ref{fig_twolayer_V0}(a) and \ref{fig_twolayer_V0}(b), compared with $\sigma_{xy}(n_z)= 0.015e^2/h$ for the trivial insulator when $E_F$ is in the band gap, for the AFM topological insulator $\sigma_{xy}(n_z)$ reaches $0.48e^2/h$, which is more than one order stronger than those in the trivial insulator as is expected.

Figure~\ref{fig_twolayer_V0}(c) shows $\sigma_{xy}\left(z\right)$ as a function of the layer index $z$ for a four-layer slab. The Hall conductance of the internal layers (i.e., $z=2$ and $z=3$) is about $\pm 0.23e^2/h$ for $m=0.12$, which in accordance with the results of the bulk (see Fig.~\ref{fig_conductance_bulk}). On the other hand, the layer Hall effect for the surface layers is much stronger than the bulk layers show in Fig.~\ref{fig_conductance_bulk}(a). This effect is attributed to the half-quantized surface effect in the axion insulator phase~\cite{Varnava18prb}(see Sec.~SIII of Supplementary file for more details).

The comparison between the trivial insulator and the AFM topological insulator can be revealed more clearly in Fig.~\ref{fig_twolayer_V0}(d), which shows the energy gap $E_g$ (dashed blue) and $\sigma_{xy}(n_z)$ (solid red) as functions of $M_0$. The topological transition happens at a larger $M_0$ with increasing $n_z$, without gap closing because of the preserved $\mathcal{PT}$ and broken $\mathcal{T}$ symmetry. In other words, there is always a mass term resulting from the interplay between the finite-size hybridization and the A-type AFM order \cite{JZhang2020CPL}. Moreover, the two layers in bilayer system are coupled and thus $\sigma_{xy}\left(n_z\right)$ is always smaller compared to the multiple-layer system.

Overall, our results reveal three distinct features to enhance the layer Hall effect in $\mathcal{PT}$ symmetric AFM insulators, (i) the direction of the AFM pattern aligns with  the geometric stacking of layers, (ii) nontrivial band topology in the bulk that manifests strong hidden Berry curvature, and (iii) the layer Hall effect is further enhanced at the surface layers in the axion insulator phase due to the surface anomalous Hall effect. In the experiment~\cite{Gao2021Nature}, Gao \emph{et al.} chose a 6-layer slab of MnBi$_2$Te$_4$ with A-type AFM pattern along the stacking $z$ direction, which fulfills all the three conditions above, and thus observed a significant layer Hall effect. More importantly, beyond the $\mathcal{PT}$-breaking physical picture in the experiment~\cite{Gao2021Nature}, we reveal a universal origin of the layer Hall effect in terms of the hidden Berry curvature physics and related design principles for future material search.

\section{Using perpendicular electric field to reveal Layer Hall effect }

\begin{figure*}[htpb]
\centering
\includegraphics[width=\linewidth]{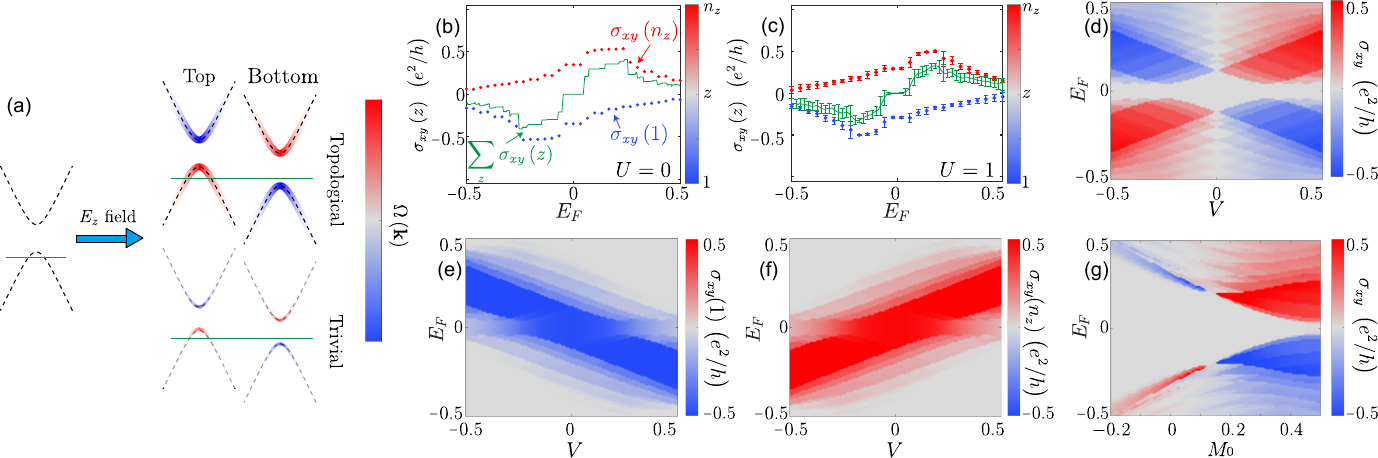}
\caption{(a) Schematic of the layer Hall effect with a perpendicular electric field. The perpendicular electric field induces a potential difference between the top and bottom surfaces, leading to an imbalance of their Hall conductances, as a result of the uncompensated hidden Berry curvature of different layers when the global $\mathcal{PT}$ symmetry is broken. (b-c) Numerically calculated Hall conductances as functions of the Fermi energy $E_F$ for AFM topological insulator thinfilms with the perpendicular electric field $V=0.3$ for the disorder strength (b) $U=0$ and (c) $U=1$. Here the dots correspond to the layer-resolved Hall conductance $\sigma_{xy}\left(z\right)$ and the color distinguishes different layer $z$. The green lines denote the net Hall conductance $\sigma_{xy}=\sum _z\sigma_{xy}\left(z\right)$. In (c), the error
bars are magnified by $3$ times to show the conductance fluctuations of 200 samples. (d-f) Numerically calculated Hall conductances $\sigma_{xy}$, $\sigma_{xy}\left(1\right)$, and $\sigma_{xy}\left(n_z\right)$ as functions of $E_F$ and $V$. (g) Numerically calculated Hall conductance $\sigma_{xy}$ as functions of $E_F$ and $M_0$. Here, we take $M_0=0.4$, $m=0.12$ in (b-f) and $V=0.3$, $m=0.12$ in (g). The system size is $n_x=n_y=40$ and $n_z=2$.  }
\label{fig_twolayer_V3}
\end{figure*}

In the bilayer system, the hidden Berry curvatures localized on each layer have degenerate energy due to the $\mathcal{PT}$ symmetry. In this section, we show how the layer-locked Berry curvature and the layer Hall effect become observable by applying a perpendicular electric field $E_z$. As shown in Fig.~\ref{fig_twolayer_V3}(a), $E_z$ breaks the $\mathcal{PT}$ symmetry and induces an energy offset between the states of the two layers. When the Fermi energy cuts the valence band of the top layer, the system is dominated by the occupied bands with the negative Berry curvature contributed by the bottom layer. As a result, a net anomalous Hall conductance appears. Here, we emphasize that the net anomalous Hall conductance is not simply originated from the $\mathcal{PT}$ breaking, but the emergence of the hidden Berry curvature. The role of $E_z$ is predominately to separate the hidden Berry curvature of different layers compensated by the global $\mathcal{PT}$ symmetry. This can be easily verified by that the Hall conductance of an AFM topological insulator is much larger than that of a trivial insulator under the same electric field [see Fig.~\ref{fig_twolayer_V3}(a)].

Figure~\ref{fig_twolayer_V3}(b) further illustrates that the imbalance of the layer-locked Berry curvatures leads to a net anomalous Hall conductance, showing two plateaus with opposite signs near $E_F=\pm V$.
The layer Hall effect is also robust against disorder [Fig.~\ref{fig_twolayer_V3}(c)]. We introduce the Anderson-type disorder to the system with $\Delta(H)=\sum_{\mathbf{r}}\phi_{\mathbf{r}}^{\dagger} W_{\mathbf{r}}\phi_{\mathbf{r}}$, where $W_{\mathbf{r}}$ is uniformly distributed within [$-U/2, U/2$], with $U$ being the disorder strength. For $U=1$, which is much larger compared to the size of the AFM gap (about 0.2), the conductance plateaus are still observable.

Figures~\ref{fig_twolayer_V3}(d-f) show the numerically calculated  $\sigma_{xy}$, $\sigma_{xy}\left(1\right)$, and $\sigma_{xy}\left(n_z\right)$ as functions of $V$ and $E_F$. With increasing field strength, the two Hall conductance plateaus of the opposite layers move along opposite directions in energy. The net anomalous Hall conductances are mainly contributed by one of the two layers~[Figs.~\ref{fig_twolayer_V3}(e-f)]. Furthermore, the chirality of the anomalous Hall effect can be efficiently tuned by modulating the direction of the perpendicular electric field. Compared to trivial insulators, the anomalous Hall conductance in AFM topological insulators is much more prominent~[Fig.~\ref{fig_twolayer_V3}(g)], which is consistent with its origin of the hidden Berry curvature, as shown in Fig.~\ref{fig_twolayer_V3}(a).


In Sec.~SII of Supplementary file, we show more results similar to those in Fig.~\ref{fig_twolayer_V3}, but for different numbers of layers, ranging from $n_z=3$ to $n_z=8$. For different even-layer films, the results are similar to the case with $n_z=2$. The odd-layer films have no $\mathcal{PT}$ symmetry, and the layer Hall effect is absent in these systems. We also show the results for different system sizes $n_{x,y}$. The Hall conductances converge as $n_{x,y}$ increase. Moreover, in Sec. SIV of Supplementary file, we show that the layer Hall effect can be revealed without breaking the $\mathcal{PT}$-symmetry, by proposing a new setup based on the nonlocal measurement.

\begin{figure}[ptb]
\centering
\includegraphics[width=1\linewidth]{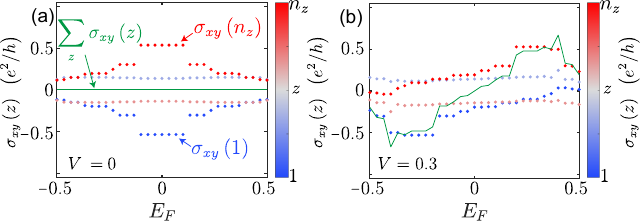}
\caption{Numerically calculated Hall conductances as functions of the Fermi energy $E_F$ for the axion insulator with (a) $V=0$ and (b) $V=0.3$. Here the dots correspond to the layer-resolved Hall conductance $\sigma_{xy}\left(z\right)$ and the color distinguishes different layer $z$. The green lines denote the net Hall conductance $\sigma_{xy}=\sum _z\sigma_{xy}\left(z\right)$.  We take $M_0=0.4$ and $m=0.12$ with the system size $n_x=n_y=40$ and $n_z=4$.}
\label{fig_fourlayer}
\end{figure}
\begin{figure*}[t]
\centering
\includegraphics[width=0.9\linewidth]{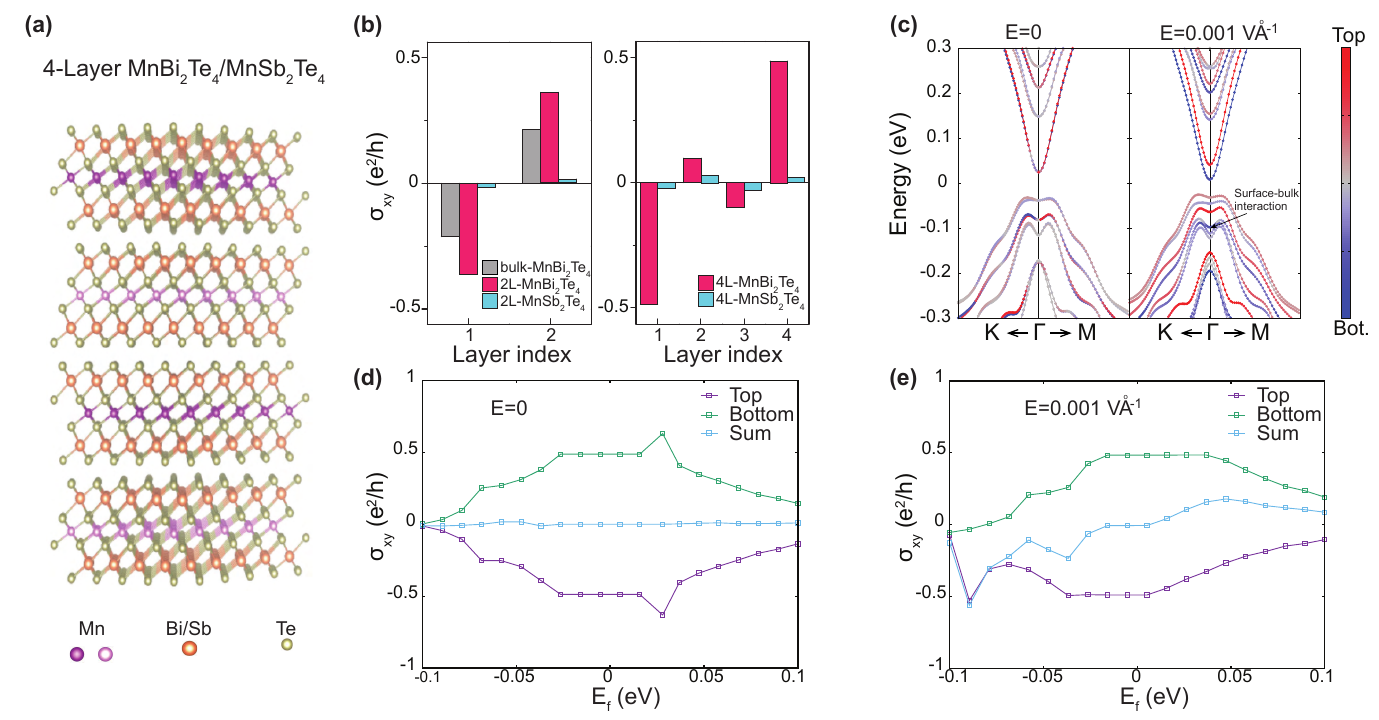}
\caption{(a) Structure of a four-layer slab of MnBi$_2$Te$_4$ or MnSb$_2$Te$_4$ computed using the first-principles theory. (b) Layer-resolved Hall conductance $\sigma_{xy}$ for  MnBi$_2$Te$_4$ (magenta) and MnSb$_2$Te$_4$ (cyan), respectively. (c) Electronic band structure for the four-layer MnBi$_2$Te$_4$ under a zero (left) and 1 mV/\AA~(right) perpendicular electric field. The red and blue dots denote the projection to the top and bottom surfaces, respectively. (d, e) The Hall conductance for the top and bottom surfaces as well as their sum for the four-layer MnBi$_2$Te$_4$ under a zero (d) and 1 mV/\AA~(e) perpendicular electric field.}
\label{fig_DFT}
\end{figure*}

\section{Layer Hall effect as a signature for axion insulator}
When increasing the film thickness $n_z$, the A-type AFM topological insulator, such as MnBi$_2$Te$_4$, exhibits the long-sought axion-insulator phase~\cite{Zhang19prl}. Such exotic phenomenon originates from the bulk topological magnetoelectric response, manifesting half-quantized surface anomalous Hall conductance $\sigma_{xy}=\pm e^2/2h$, which could serve as an experimental evidence. However, up to now the verification of the axion insulator is limited to the effect of zero Hall plateau, which is an indirect evidence for the axion insulator~\cite{Wang15prbrc,Liu20nm,Mogi2017NatMat}. In this section, we elucidate that the layer Hall effect is an ideal physical observable that is feasible to measure, providing a signature for the axion-insulator phase~\cite{Gu2020Arxiv}.

As shown in Fig.~\ref{fig_twolayer_V0}(c), $\sigma_{xy}(n_z)$ converges at $M_0=0.4$ when $n_z \ge 4$, indicating that a four-layer system is enough to present the surface effect of a three-dimensional axion-insulator phase. Figures~\ref{fig_fourlayer}(a-b) show the numerically calculated Hall conductances as functions of the Fermi energy $E_F$ for the four-layer AFM topological insulator under the electric field with $V=0$ and $V=0.3$, respectively. We find that the enhancement of the layer Hall effect mainly reflects on the surface layers [Fig.~\ref{fig_fourlayer}(a)]. This still works for the case with larger $n_z$~(see Supplementary file). In comparison, the interior layers contribute weaker layer Hall conductance (about $\pm 0.13 e^2/h$ inside the band gap) due to the oscillation of the layer-dependent magnetization~\cite{Gu2020Arxiv}.

The axion insulator is characterized by two intertwined effects, one is the half-quantized surface Hall conductance, the other is the bulk magnetoelectric effect. Therefore, the observation of the layer Hall effect characterized by the nearly half-quantized Hall conductance implies the possibility of the axion insulator phase and its related the bulk topological magnetoelectric effect. On the other hand, the layer Hall effect has both the surface and bulk contributions. In the axion insulator, the surface Hall conductance is oppositely half-quantized on the top and bottom surfaces, because of the bulk magnetoelectric coupling. Due to the contribution from the bulk, the layer Hall conductance deviates from the precise half-quantization, but its large signal in the experiments could provide hints for the co-existing half-quantized surfaces and possible axion insulator phase.
Remarkably, in the presence of the perpendicular electric field, the two anomalous Hall conductance plateaus shown in Fig.~\ref{fig_fourlayer}(b) reach nearly half-quantized values $\pm e^2/2h$ when the energy degeneracy of the two surfaces are lifted. Despite that the half quantization of the surface anomalous Hall effect is not topologically protected in a metallic state, the nearly $\pm e^2/2h$ Hall conductance, the sign of which depends on the field direction, serving as a signature of axion insulator phase and its related the bulk topological magnetoelectric effect. Therefore, we suggest that the layer Hall effect can provide a signature for detecting the long-sought axion-insulator phase.


\section{$\text{MnSb}_2\text{Te}_4$ and $\text{MnBi}_2\text{Te}_4$ from first-principles calculations}

To further verify the layer Hall effect, we next perform first-principles calculations on MnBi$_2$Te$_4$, which is an ideal material platform owing to its intrinsic A-type AFM configuration [Fig.~\ref{fig_DFT}(a)]~\cite{Otrokov19nat,Deng20sci}. The first-principles calculations were performed with the Vienna \emph{Ab-Initio} Simulation Package \cite{Kresse/Furthmuller:1996a, Kresse/Joubert:1999} and the projector augmented wave  method \cite{Blochl:1994} to treat the core and valence electrons using the following  electronic configurations:
$3p^64s^23d^5$ for Mn, $5d^{10}6s^26p^3$ for Bi, $4d^{10}5s^25p^3$ for Sb, and $5s^25p^4$ for Te. The revised Perdew-Burke-Ernzerhof exchange-correlation functional \cite{PBE} was selected. The Brillouin zone is sampled using an $6\times6\times3$ $\Gamma$-centered Monkhorst-Pack $k$-point mesh and integrations are performed using Gaussian smearing with a width of 2\,meV.
We use the Perdew-Burke-Ernzerhof plus Hubbard $U$ method of Dudarev \emph{et al.} \cite{Dudarev/Zuo_et_al:2506} with $U_\mathrm{eff}(\mathrm {Mn})=5$\,eV to take into account the correlation effects of the Mn-$3d$ electrons.

Here, both the bilayer and four-layer MnBi$_2$Te$_4$ slabs are calculated. In MnBi$_2$Te$_4$, the penetration depth of the surface states is mainly embedded in the first two septuple layers from the terminating surface~\cite{Sun20prbrc,Shikin2020scirep}. Therefore, the top and bottom surface states in bilayer system are always coupled and thus cannot host the half-quantized anomalous Hall conductance, which is the manifestation of an axion insulator defined in 3D. On the other hand, a four-layer slab is just thick enough to capture the essential topology of the axion state in MnBi$_2$Te$_4$ due to its negligible finite-size effect. After projecting the Bl\"{o}ch orbitals to the Wannier orbitals, we calculate the layer-resolved Hall conductance by calculating the local Chern marker. Its calculation for a particular layer follows the derivation by Varnava \emph{et al.}~\cite{Varnava18prb}, whereas the position operators $x(y)_=\frac{\langle \psi_{ik}|i\hbar v_{x(y)}|\psi_{jk}\rangle}{E_{ik}-E_{jk}}$ are obtained within the wannier orbitals constructed by the wannier90 package interfaced to the Vienna \emph{Ab-Initio} Simulation Package~\cite{wannier90,wanniertools}. We find that the calculated $\sigma_{xy}(z)$ for the bilayer MnBi$_2$Te$_4$ reaches $\pm 0.35$ $e^2/h$. For the four-layer slab, the calculated $\sigma_{xy}\left(z\right)$ of the top and bottom layers are nearly half-quantized, i.e., $\pm0.49$ $e^2/h$, while the internal layers show much smaller contributions~[Fig.~\ref{fig_DFT}(b)].
Comparing the layer-resolved Hall conductance for the two-layer and four-layer slabs to that for the bulk MnBi$_2$Te$_4$, $\sigma_{xy}\sim\pm 0.21$ $e^2/h$ showed with the grey color bar in left panel of Fig.~\ref{fig_DFT}(b), one can clearly distinguish the contribution of Hall conductance from surface enhancement and that from strong hidden Berry curvature.

By shifting the Fermi energy, the Hall conductance deviates from $0.5$ $e^2/h$ when it cuts the valence or conduction bands~[Fig.~\ref{fig_DFT}(d)], with the summation of the top and bottom layers being zero, which is regulated by inversion symmetry. When a perpendicular electric field of $1$ mV/\AA~is applied, the global $\mathcal{PT}$  symmetry is broken and the degeneracy of the top and bottom surfaces is lifted~[Fig.~\ref{fig_DFT}(c)]. Hence, the layer Hall effect is revealed as the total Hall conductance when the Fermi energy cuts the bands of one surface, with the sign being switchable by changing the field direction.

We also calculate the layer Hall effect of a four-layer MnBi$_2$Te$_4$ slab with in-plane magnetic moments, and find that the layer Hall conductance is about $0.003$ $e^2/h$, two orders smaller than that with the out-of-plane magnetic moments. This result is consistent with the model calculations shown in Fig.~\ref{fig_conductance_bulk}(c). Although the in-plane AFM configuration also holds the global but breaks the local $\mathcal{PT}$ symmetry, there are two main differences compared to the out-of-plane configuration. Firstly, as discussed in Fig.~\ref{fig_conductance_bulk}(c), the dominant local Hall conductance for the in-plane magnetic moment case $(m//x)$ is $\sigma_{yz}$ rather than $\sigma_{xy}$, while the layer Hall effect in 2D slabs is determined by $\sigma_{xy}$. Secondly, the in-plane moment is not sufficient to open an exchange gap on the surface~\cite{prx.9.041038}. As a result, the gap opened for the four-layer slab is caused by the quantum confinement, and thus the layer Hall effect cannot be enhanced topologically.

In order to highlight the significant enhancement of the layer Hall effect in AFM topological insulators, we also perform the same calculations on a trivial insulator MnSb$_2$Te$_4$, for which the band gap is close to that in MnBi$_2$Te$_4$. The only difference in terms of the electronic structure between these two materials is that the latter has band inversion. However, the existence of the band inversion leads to a remarkable difference in the Hall conductance of the surface layers. One can find in Fig.~\ref{fig_DFT}(b) that for no matter 2-layer or 4-layer system, the Hall conductance for the top and bottom layers of MnSb$_2$Te$_4$ is one order of magnitude smaller than that of MnBi$_2$Te$_4$. The layer-locked Berry curvature of the interior layers of 4-layer MnSb$_2$Te$_4$ is slightly larger than those of the surface layers, but still much smaller than those of MnBi$_2$Te$_4$.

%

Based on the atomistic Hamiltonians with close reliance on realistic attributes of materials, our first-principles calculations yield qualitatively consistent results  with the tight-binding model calculation, while several subtle details are noted. First,
we find that under the electric field, for n-doping the total Hall conductance cannot reach $0.5e^2/h$ before the Fermi energy cuts the second conduction band [denoted by the green region in Figs.~\ref{fig_DFT}(c) and \ref{fig_DFT}(e)], when the total Hall conductance starts to drop.
Second, there is no electron-hole symmetry of the valence and conduction bands in such realistic systems. Therefore, the behavior of the layer Hall effect for p- and n-doping should be different in the experiment.
We find that the total Hall conductance for MnBi$_2$Te$_4$ can reach up to beyond $-0.5$ $e^2/h$ when $E_F$ approaches $-0.08$ eV.
From the layer-resolved Berry curvature contribution, one can find that such significant increase is due to the peak from bottom surface, which originates from the subtle interplay between the surface layer and the internal layers, denoted by the arrow in Fig.~\ref{fig_DFT}(c).

To further demonstrate the generic nature of the layer Hall effect, we also performed DFT calculations for Mn$_2$Bi$_2$Te$_5$, in which the neighboring Mn atoms are connected by the Mn-Te-Mn bond. The computed $\sigma_{xy}^{\text{LHE}}\sim 0.27 e^2/h$ is of a similar magnitude as that for MnBi$_2$Te$_4$ (see Sec.~SV of Supplementary file for more details). Therefore, we expect that the layer Hall effect should also exist in non-van der Waals materials~\cite{Xu2019PRL,acscentsci.9b00202,adma.201907565}.
Moreover, we expect that the layer Hall effect would emerge by stacking two AFM non-collinear layers~\cite{PhysRevLett.112.017205,Nakatsuji_2015} in a $\mathcal{PT}$-symmetric fashion, which might trigger further exploration in the future.

\section{Conclusion}
To summarize, we show that electron accumulation could take place at opposite edges at different layers in the layered structures with globally-preserved but locally-broken $\mathcal{PT}$ symmetry. Such a layer Hall effect exhibits a macroscopic Hall conductance by applying a perpendicular electric field, which lifts the degeneracy and thus reveals the hidden Berry curvature localized in specific layers. The recent experiment~\cite{Gao2021Nature} has demonstrated the layer Hall effect in the magnetic axion insulator MnBi$_2$Te$_4$, in terms of the anomalous Hall conductance as a perpendicular electric field breaks the $\mathcal{PT}$-symmetry. Beyond that, our theory has revealed a universal origin of the layer Hall effect in terms of the distribution and compensation of real-space-resolved Berry curvature. Such a mechanism of hidden physics not only can be generalized to many scenarios but also indicates that the layer Hall effect could be detected without breaking the global $\mathcal{PT}$-symmetry, e.g., in the nonlocal measurement (see Sec.~SIV of Supplementary file). Moreover, we have proposed more material candidates in general AFM insulators based on the layer degree of freedom, and shown why the layer Hall effect can be significantly enhanced in AFM topological insulators. This will inspire more experimental explorations.

\appendix

\section{Acknowledgments}
We thank helpful discussions with Su-Yang Xu, Bo Fu, Chui-Zhen Chen, Dong-Hui Xu, and Bin Zhou. The numerical calculations were supported by Center for Computational Science and Engineering of SUSTech.

\section{FUNDING}
This work was supported by the National Key R\&D Program of China (Grant No. 2020YFA0308900), the National Natural Science Foundation of China (11925402,11874195), the National Basic Research Program of China (2015CB921102), the Strategic Priority Research Program of the Chinese Academy of Sciences (XDB28000000), the Basic Science Center Project of NSFC (51788104), Guangdong College Innovation Team (2016ZT06D348, 2020KCXTD001), Shenzhen High-level Special Fund (G02206304, G02206404), Guangdong Provincial Key Laboratory for Computational Science and Material Design (Grant No. 2019B030301001), and the Science, Technology and Innovation Commission of Shenzhen Municipality (ZDSYS20170303165926217, JCYJ20170412152620376, KYTDPT20181011104202253). R.C. acknowledges support from the project funded by the China Postdoctoral Science Foundation (Grant No. 2019M661678) and the SUSTech Presidential Postdoctoral Fellowship. M.G. was supported by Natural Science Foundation of Guangdong Province (2021A1515110389), Science Technology and Innovation Commission of Shenzhen Municipality (JCYJ20210324104812034), and Foundation for Distinguished Young Talents in Higher Education of Guangdong Province (2020KQNCX064).

\bibliographystyle{apsrev4-1-etal-title_6authors}
\bibliography{refs-transport,refs-transport-chen,bibfile,bibfile_dft}

\begin{thebibliography}{50}%
\makeatletter
\providecommand \@ifxundefined [1]{%
 \@ifx{#1\undefined}
}%
\providecommand \@ifnum [1]{%
 \ifnum #1\expandafter \@firstoftwo
 \else \expandafter \@secondoftwo
 \fi
}%
\providecommand \@ifx [1]{%
 \ifx #1\expandafter \@firstoftwo
 \else \expandafter \@secondoftwo
 \fi
}%
\providecommand \natexlab [1]{#1}%
\providecommand \enquote  [1]{``#1''}%
\providecommand \bibnamefont  [1]{#1}%
\providecommand \bibfnamefont [1]{#1}%
\providecommand \citenamefont [1]{#1}%
\providecommand \href@noop [0]{\@secondoftwo}%
\providecommand \href [0]{\begingroup \@sanitize@url \@href}%
\providecommand \@href[1]{\@@startlink{#1}\@@href}%
\providecommand \@@href[1]{\endgroup#1\@@endlink}%
\providecommand \@sanitize@url [0]{\catcode `\\12\catcode `\$12\catcode
  `\&12\catcode `\#12\catcode `\^12\catcode `\_12\catcode `\%12\relax}%
\providecommand \@@startlink[1]{}%
\providecommand \@@endlink[0]{}%
\providecommand \url  [0]{\begingroup\@sanitize@url \@url }%
\providecommand \@url [1]{\endgroup\@href {#1}{\urlprefix }}%
\providecommand \urlprefix  [0]{URL }%
\providecommand \Eprint [0]{\href }%
\providecommand \doibase [0]{http://dx.doi.org/}%
\providecommand \selectlanguage [0]{\@gobble}%
\providecommand \bibinfo  [0]{\@secondoftwo}%
\providecommand \bibfield  [0]{\@secondoftwo}%
\providecommand \translation [1]{[#1]}%
\providecommand \BibitemOpen [0]{}%
\providecommand \bibitemStop [0]{}%
\providecommand \bibitemNoStop [0]{.\EOS\space}%
\providecommand \EOS [0]{\spacefactor3000\relax}%
\providecommand \BibitemShut  [1]{\csname bibitem#1\endcsname}%
\let\auto@bib@innerbib\@empty
\bibitem [{\citenamefont {Nagaosa}\ \emph {et~al.}(2010)\citenamefont
  {Nagaosa}, \citenamefont {Sinova}, \citenamefont {Onoda}, \citenamefont
  {MacDonald},\ and\ \citenamefont {Ong}}]{Nagaosa10rmp}%
  \BibitemOpen
  \bibfield  {author} {\bibinfo {author} {\bibfnamefont {N.}~\bibnamefont
  {Nagaosa}}, \bibinfo {author} {\bibfnamefont {J.}~\bibnamefont {Sinova}},
  \bibinfo {author} {\bibfnamefont {S.}~\bibnamefont {Onoda}}, \bibinfo
  {author} {\bibfnamefont {A.~H.}\ \bibnamefont {MacDonald}}, \ and\ \bibinfo
  {author} {\bibfnamefont {N.~P.}\ \bibnamefont {Ong}},\ }\bibfield  {title}
  {\enquote {\bibinfo {title} {Anomalous {Hall} effect}}, }\href {\doibase
  10.1103/RevModPhys.82.1539} {\bibfield  {journal} {\bibinfo  {journal} {Rev.
  Mod. Phys.}\ }\textbf {\bibinfo {volume} {82}},\ \bibinfo {pages} {1539}
  (\bibinfo {year} {2010})}\BibitemShut {NoStop}%
\bibitem [{\citenamefont {Sinova}\ \emph {et~al.}(2015)\citenamefont {Sinova},
  \citenamefont {Valenzuela}, \citenamefont {Wunderlich}, \citenamefont
  {Back},\ and\ \citenamefont {Jungwirth}}]{Sinova15rmp}%
  \BibitemOpen
  \bibfield  {author} {\bibinfo {author} {\bibfnamefont {J.}~\bibnamefont
  {Sinova}}, \bibinfo {author} {\bibfnamefont {S.~O.}\ \bibnamefont
  {Valenzuela}}, \bibinfo {author} {\bibfnamefont {J.}~\bibnamefont
  {Wunderlich}}, \bibinfo {author} {\bibfnamefont {C.~H.}\ \bibnamefont
  {Back}}, \ and\ \bibinfo {author} {\bibfnamefont {T.}~\bibnamefont
  {Jungwirth}},\ }\bibfield  {title} {\enquote {\bibinfo {title} {Spin {Hall}
  effects}}, }\href {\doibase 10.1103/RevModPhys.87.1213} {\bibfield  {journal}
  {\bibinfo  {journal} {Rev. Mod. Phys.}\ }\textbf {\bibinfo {volume} {87}},\
  \bibinfo {pages} {1213} (\bibinfo {year} {2015})}\BibitemShut {NoStop}%
\bibitem [{\citenamefont {Bernevig}\ \emph {et~al.}(2006)\citenamefont
  {Bernevig}, \citenamefont {Hughes},\ and\ \citenamefont
  {Zhang}}]{Bernevig06sci}%
  \BibitemOpen
  \bibfield  {author} {\bibinfo {author} {\bibfnamefont {B.~A.}\ \bibnamefont
  {Bernevig}}, \bibinfo {author} {\bibfnamefont {T.~L.}\ \bibnamefont
  {Hughes}}, \ and\ \bibinfo {author} {\bibfnamefont {S.-C.}\ \bibnamefont
  {Zhang}},\ }\bibfield  {title} {\enquote {\bibinfo {title} {Quantum spin
  {Hall} effect and topological phase transition in {HgTe} quantum wells}},
  }\href {http://science.sciencemag.org/content/314/5806/1757} {\bibfield
  {journal} {\bibinfo  {journal} {Science}\ }\textbf {\bibinfo {volume}
  {314}},\ \bibinfo {pages} {1757} (\bibinfo {year} {2006})}\BibitemShut
  {NoStop}%
\bibitem [{\citenamefont {Xiao}\ \emph {et~al.}(2007)\citenamefont {Xiao},
  \citenamefont {Yao},\ and\ \citenamefont {Niu}}]{Xiao07prl}%
  \BibitemOpen
  \bibfield  {author} {\bibinfo {author} {\bibfnamefont {D.}~\bibnamefont
  {Xiao}}, \bibinfo {author} {\bibfnamefont {W.}~\bibnamefont {Yao}}, \ and\
  \bibinfo {author} {\bibfnamefont {Q.}~\bibnamefont {Niu}},\ }\bibfield
  {title} {\enquote {\bibinfo {title} {{Valley-contrasting physics in graphene:
  magnetic moment and topological transport}}}, }\href {\doibase
  10.1103/PhysRevLett.99.236809} {\bibfield  {journal} {\bibinfo  {journal}
  {Phys. Rev. Lett.}\ }\textbf {\bibinfo {volume} {99}},\ \bibinfo {pages}
  {236809} (\bibinfo {year} {2007})}\BibitemShut {NoStop}%
\bibitem [{\citenamefont {Kato}\ \emph {et~al.}(2004)\citenamefont {Kato},
  \citenamefont {Myers}, \citenamefont {Gossard},\ and\ \citenamefont
  {Awschalom}}]{Kato04sci}%
  \BibitemOpen
  \bibfield  {author} {\bibinfo {author} {\bibfnamefont {Y.~K.}\ \bibnamefont
  {Kato}}, \bibinfo {author} {\bibfnamefont {R.~C.}\ \bibnamefont {Myers}},
  \bibinfo {author} {\bibfnamefont {A.~C.}\ \bibnamefont {Gossard}}, \ and\
  \bibinfo {author} {\bibfnamefont {D.~D.}\ \bibnamefont {Awschalom}},\
  }\bibfield  {title} {\enquote {\bibinfo {title} {{Observation of the spin
  Hall effect in semiconductors}}}, }\href {\doibase 10.1126/science.1105514}
  {\bibfield  {journal} {\bibinfo  {journal} {Science}\ }\textbf {\bibinfo
  {volume} {306}},\ \bibinfo {pages} {1910} (\bibinfo {year}
  {2004})}\BibitemShut {NoStop}%
\bibitem [{\citenamefont {K{\"o}nig}\ \emph {et~al.}(2007)\citenamefont
  {K{\"o}nig}, \citenamefont {Wiedmann}, \citenamefont {Br{\"u}ne},
  \citenamefont {Roth}, \citenamefont {Buhmann}, \citenamefont {Molenkamp},
  \citenamefont {Qi},\ and\ \citenamefont {Zhang}}]{Konig07sci}%
  \BibitemOpen
  \bibfield  {author} {\bibinfo {author} {\bibfnamefont {M.}~\bibnamefont
  {K{\"o}nig}}, \bibinfo {author} {\bibfnamefont {S.}~\bibnamefont {Wiedmann}},
  \bibinfo {author} {\bibfnamefont {C.}~\bibnamefont {Br{\"u}ne}}, \bibinfo
  {author} {\bibfnamefont {A.}~\bibnamefont {Roth}}, \bibinfo {author}
  {\bibfnamefont {H.}~\bibnamefont {Buhmann}}, \bibinfo {author} {\bibfnamefont
  {L.~W.}\ \bibnamefont {Molenkamp}}, \bibinfo {author} {\bibfnamefont {X.-L.}\
  \bibnamefont {Qi}}, \ and\ \bibinfo {author} {\bibfnamefont {S.-C.}\
  \bibnamefont {Zhang}},\ }\bibfield  {title} {\enquote {\bibinfo {title}
  {Quantum spin {Hall} insulator state in {HgTe} quantum wells}}, }\href
  {\doibase 10.1126/science.1148047} {\bibfield  {journal} {\bibinfo  {journal}
  {Science}\ }\textbf {\bibinfo {volume} {318}},\ \bibinfo {pages} {766}
  (\bibinfo {year} {2007})}\BibitemShut {NoStop}%
\bibitem [{\citenamefont {Roth}\ \emph {et~al.}(2009)\citenamefont {Roth},
  \citenamefont {Brune}, \citenamefont {Buhmann}, \citenamefont {Molenkamp},
  \citenamefont {Maciejko}, \citenamefont {Qi},\ and\ \citenamefont
  {Zhang}}]{Roth09sci}%
  \BibitemOpen
  \bibfield  {author} {\bibinfo {author} {\bibfnamefont {A.}~\bibnamefont
  {Roth}}, \bibinfo {author} {\bibfnamefont {C.}~\bibnamefont {Brune}},
  \bibinfo {author} {\bibfnamefont {H.}~\bibnamefont {Buhmann}}, \bibinfo
  {author} {\bibfnamefont {L.~W.}\ \bibnamefont {Molenkamp}}, \bibinfo {author}
  {\bibfnamefont {J.}~\bibnamefont {Maciejko}}, \bibinfo {author}
  {\bibfnamefont {X.-L.}\ \bibnamefont {Qi}}, \ and\ \bibinfo {author}
  {\bibfnamefont {S.-C.}\ \bibnamefont {Zhang}},\ }\bibfield  {title} {\enquote
  {\bibinfo {title} {Nonlocal transport in the quantum spin {Hall} state}},
  }\href {\doibase 10.1126/science.1174736} {\bibfield  {journal} {\bibinfo
  {journal} {Science}\ }\textbf {\bibinfo {volume} {325}},\ \bibinfo {pages}
  {294} (\bibinfo {year} {2009})}\BibitemShut {NoStop}%
\bibitem [{\citenamefont {Mak}\ \emph {et~al.}(2014)\citenamefont {Mak},
  \citenamefont {McGill}, \citenamefont {Park},\ and\ \citenamefont
  {McEuen}}]{Mak14sci}%
  \BibitemOpen
  \bibfield  {author} {\bibinfo {author} {\bibfnamefont {K.~F.}\ \bibnamefont
  {Mak}}, \bibinfo {author} {\bibfnamefont {K.~L.}\ \bibnamefont {McGill}},
  \bibinfo {author} {\bibfnamefont {J.}~\bibnamefont {Park}}, \ and\ \bibinfo
  {author} {\bibfnamefont {P.~L.}\ \bibnamefont {McEuen}},\ }\bibfield  {title}
  {\enquote {\bibinfo {title} {The valley {Hall} effect in {MoS}$_2$
  transistors}}, }\href {\doibase 10.1126/science.1250140} {\bibfield
  {journal} {\bibinfo  {journal} {Science}\ }\textbf {\bibinfo {volume}
  {344}},\ \bibinfo {pages} {1489} (\bibinfo {year} {2014})}\BibitemShut
  {NoStop}%
\bibitem [{\citenamefont {Chang}\ \emph {et~al.}(2013)\citenamefont {Chang},
  \citenamefont {Zhang}, \citenamefont {Feng}, \citenamefont {Shen},
  \citenamefont {Zhang}, \citenamefont {Guo}, \citenamefont {Li}, \citenamefont
  {Ou}, \citenamefont {Wei}, \citenamefont {Wang}, \citenamefont {Ji},
  \citenamefont {Feng}, \citenamefont {Ji}, \citenamefont {Chen}, \citenamefont
  {Jia}, \citenamefont {Dai}, \citenamefont {Fang}, \citenamefont {Zhang},
  \citenamefont {He}, \citenamefont {Wang}, \citenamefont {Lu}, \citenamefont
  {Ma},\ and\ \citenamefont {Xue}}]{Chang13sci}%
  \BibitemOpen
  \bibfield  {author} {\bibinfo {author} {\bibfnamefont {C.-Z.}\ \bibnamefont
  {Chang}}, \bibinfo {author} {\bibfnamefont {J.}~\bibnamefont {Zhang}},
  \bibinfo {author} {\bibfnamefont {X.}~\bibnamefont {Feng}}, \bibinfo {author}
  {\bibfnamefont {J.}~\bibnamefont {Shen}}, \bibinfo {author} {\bibfnamefont
  {Z.}~\bibnamefont {Zhang}}, \bibinfo {author} {\bibfnamefont
  {M.}~\bibnamefont {Guo}},  \emph {et~al.},\ }\bibfield  {title} {\enquote
  {\bibinfo {title} {Experimental observation of the quantum anomalous {Hall}
  effect in a magnetic topological insulator}}, }\href {\doibase
  10.1126/science.1234414} {\bibfield  {journal} {\bibinfo  {journal}
  {Science}\ }\textbf {\bibinfo {volume} {340}},\ \bibinfo {pages} {167}
  (\bibinfo {year} {2013})}\BibitemShut {NoStop}%
\bibitem [{\citenamefont {Gao}\ \emph {et~al.}(2021)\citenamefont {Gao},
  \citenamefont {Liu}, \citenamefont {Hu}, \citenamefont {Qiu}, \citenamefont
  {Tzschaschel}, \citenamefont {Ghosh}, \citenamefont {Ho}, \citenamefont
  {B{\'{e}}rub{\'{e}}}, \citenamefont {Chen}, \citenamefont {Sun},
  \citenamefont {Zhang}, \citenamefont {Zhang}, \citenamefont {Wang},
  \citenamefont {Wang}, \citenamefont {Huang}, \citenamefont {Felser},
  \citenamefont {Agarwal}, \citenamefont {Ding}, \citenamefont {Tien},
  \citenamefont {Akey}, \citenamefont {Gardener}, \citenamefont {Singh},
  \citenamefont {Watanabe}, \citenamefont {Taniguchi}, \citenamefont {Burch},
  \citenamefont {Bell}, \citenamefont {Zhou}, \citenamefont {Gao},
  \citenamefont {Lu}, \citenamefont {Bansil}, \citenamefont {Lin},
  \citenamefont {Chang}, \citenamefont {Fu}, \citenamefont {Ma}, \citenamefont
  {Ni},\ and\ \citenamefont {Xu}}]{Gao2021Nature}%
  \BibitemOpen
  \bibfield  {author} {\bibinfo {author} {\bibfnamefont {A.}~\bibnamefont
  {Gao}}, \bibinfo {author} {\bibfnamefont {Y.-F.}\ \bibnamefont {Liu}},
  \bibinfo {author} {\bibfnamefont {C.}~\bibnamefont {Hu}}, \bibinfo {author}
  {\bibfnamefont {J.-X.}\ \bibnamefont {Qiu}}, \bibinfo {author} {\bibfnamefont
  {C.}~\bibnamefont {Tzschaschel}}, \bibinfo {author} {\bibfnamefont
  {B.}~\bibnamefont {Ghosh}},  \emph {et~al.},\ }\bibfield  {title} {\enquote
  {\bibinfo {title} {{Layer Hall effect in a 2D topological axion
  antiferromagnet}}}, }\href {\doibase 10.1038/s41586-021-03679-w} {\bibfield
  {journal} {\bibinfo  {journal} {Nature}\ }\textbf {\bibinfo {volume} {595}},\
  \bibinfo {pages} {521} (\bibinfo {year} {2021})}\BibitemShut {NoStop}%
\bibitem [{\citenamefont {Dai}\ \emph {et~al.}(2022)\citenamefont {Dai},
  \citenamefont {Li}, \citenamefont {Xu}, \citenamefont {Chen},\ and\
  \citenamefont {Xie}}]{ChenCZarXiv2022}%
  \BibitemOpen
  \bibfield  {author} {\bibinfo {author} {\bibfnamefont {W.-B.}\ \bibnamefont
  {Dai}}, \bibinfo {author} {\bibfnamefont {H.}~\bibnamefont {Li}}, \bibinfo
  {author} {\bibfnamefont {D.-H.}\ \bibnamefont {Xu}}, \bibinfo {author}
  {\bibfnamefont {C.-Z.}\ \bibnamefont {Chen}}, \ and\ \bibinfo {author}
  {\bibfnamefont {X.}~\bibnamefont {Xie}},\ }\bibfield  {title} {\enquote
  {\bibinfo {title} {{Quantum anomalous layer Hall effect in the topological
  magnet MnBi$_2$Te$_4$}}}, }\href@noop {} {\ ,\ \bibinfo {pages}
  {arXiv:2206.09635} (\bibinfo {year} {2022})}\BibitemShut {NoStop}%
\bibitem [{\citenamefont {Otrokov}\ \emph {et~al.}(2019)\citenamefont
  {Otrokov}, \citenamefont {Klimovskikh}, \citenamefont {Bentmann},
  \citenamefont {Estyunin}, \citenamefont {Zeugner}, \citenamefont {Aliev},
  \citenamefont {Ga{\ss}}, \citenamefont {Wolter}, \citenamefont {Koroleva},
  \citenamefont {Shikin}, \citenamefont {Blanco-Rey}, \citenamefont {Hoffmann},
  \citenamefont {Rusinov}, \citenamefont {Vyazovskaya}, \citenamefont
  {Eremeev}, \citenamefont {Koroteev}, \citenamefont {Kuznetsov}, \citenamefont
  {Freyse}, \citenamefont {S{\'a}nchez-Barriga}, \citenamefont {Amiraslanov},
  \citenamefont {Babanly}, \citenamefont {Mamedov}, \citenamefont {Abdullayev},
  \citenamefont {Zverev}, \citenamefont {Alfonsov}, \citenamefont {Kataev},
  \citenamefont {B{\"u}chner}, \citenamefont {Schwier}, \citenamefont {Kumar},
  \citenamefont {Kimura}, \citenamefont {Petaccia}, \citenamefont {Di~Santo},
  \citenamefont {Vidal}, \citenamefont {Schatz}, \citenamefont {Ki{\ss}ner},
  \citenamefont {{\"U}nzelmann}, \citenamefont {Min}, \citenamefont {Moser},
  \citenamefont {Peixoto}, \citenamefont {Reinert}, \citenamefont {Ernst},
  \citenamefont {Echenique}, \citenamefont {Isaeva},\ and\ \citenamefont
  {Chulkov}}]{Otrokov19nat}%
  \BibitemOpen
  \bibfield  {author} {\bibinfo {author} {\bibfnamefont {M.~M.}\ \bibnamefont
  {Otrokov}}, \bibinfo {author} {\bibfnamefont {I.~I.}\ \bibnamefont
  {Klimovskikh}}, \bibinfo {author} {\bibfnamefont {H.}~\bibnamefont
  {Bentmann}}, \bibinfo {author} {\bibfnamefont {D.}~\bibnamefont {Estyunin}},
  \bibinfo {author} {\bibfnamefont {A.}~\bibnamefont {Zeugner}}, \bibinfo
  {author} {\bibfnamefont {Z.~S.}\ \bibnamefont {Aliev}},  \emph {et~al.},\
  }\bibfield  {title} {\enquote {\bibinfo {title} {Prediction and observation
  of an antiferromagnetic topological insulator}}, }\href
  {https://doi.org/10.1038/s41586-019-1840-9} {\bibfield  {journal} {\bibinfo
  {journal} {Nature}\ }\textbf {\bibinfo {volume} {576}},\ \bibinfo {pages}
  {416} (\bibinfo {year} {2019})}\BibitemShut {NoStop}%
\bibitem [{\citenamefont {Deng}\ \emph {et~al.}(2020)\citenamefont {Deng},
  \citenamefont {Yu}, \citenamefont {Shi}, \citenamefont {Guo}, \citenamefont
  {Xu}, \citenamefont {Wang}, \citenamefont {Chen},\ and\ \citenamefont
  {Zhang}}]{Deng20sci}%
  \BibitemOpen
  \bibfield  {author} {\bibinfo {author} {\bibfnamefont {Y.}~\bibnamefont
  {Deng}}, \bibinfo {author} {\bibfnamefont {Y.}~\bibnamefont {Yu}}, \bibinfo
  {author} {\bibfnamefont {M.~Z.}\ \bibnamefont {Shi}}, \bibinfo {author}
  {\bibfnamefont {Z.}~\bibnamefont {Guo}}, \bibinfo {author} {\bibfnamefont
  {Z.}~\bibnamefont {Xu}}, \bibinfo {author} {\bibfnamefont {J.}~\bibnamefont
  {Wang}}, \bibinfo {author} {\bibfnamefont {X.~H.}\ \bibnamefont {Chen}}, \
  and\ \bibinfo {author} {\bibfnamefont {Y.}~\bibnamefont {Zhang}},\ }\bibfield
   {title} {\enquote {\bibinfo {title} {Quantum anomalous {Hall} effect in
  intrinsic magnetic topological insulator {MnBi}$_2${Te}$_4$}}, }\href
  {https://science.sciencemag.org/content/367/6480/895} {\bibfield  {journal}
  {\bibinfo  {journal} {Science}\ }\textbf {\bibinfo {volume} {367}},\ \bibinfo
  {pages} {895} (\bibinfo {year} {2020})}\BibitemShut {NoStop}%
\bibitem [{\citenamefont {Chen}\ \emph {et~al.}(2019)\citenamefont {Chen},
  \citenamefont {Liu},\ and\ \citenamefont {Xie}}]{ChenCZ2019PRL}%
  \BibitemOpen
  \bibfield  {author} {\bibinfo {author} {\bibfnamefont {C.-Z.}\ \bibnamefont
  {Chen}}, \bibinfo {author} {\bibfnamefont {H.}~\bibnamefont {Liu}}, \ and\
  \bibinfo {author} {\bibfnamefont {X.~C.}\ \bibnamefont {Xie}},\ }\bibfield
  {title} {\enquote {\bibinfo {title} {{Effects of random domains on the zero
  Hall plateau in the quantum anomalous Hall effect}}}, }\href {\doibase
  10.1103/PhysRevLett.122.026601} {\bibfield  {journal} {\bibinfo  {journal}
  {Phys. Rev. Lett.}\ }\textbf {\bibinfo {volume} {122}},\ \bibinfo {pages}
  {026601} (\bibinfo {year} {2019})}\BibitemShut {NoStop}%
\bibitem [{\citenamefont {Chen}\ \emph
  {et~al.}(2021{\natexlab{a}})\citenamefont {Chen}, \citenamefont {Li},
  \citenamefont {Sun}, \citenamefont {Liu}, \citenamefont {Zhao}, \citenamefont
  {Lu},\ and\ \citenamefont {Xie}}]{ChenR2020arXiv}%
  \BibitemOpen
  \bibfield  {author} {\bibinfo {author} {\bibfnamefont {R.}~\bibnamefont
  {Chen}}, \bibinfo {author} {\bibfnamefont {S.}~\bibnamefont {Li}}, \bibinfo
  {author} {\bibfnamefont {H.-P.}\ \bibnamefont {Sun}}, \bibinfo {author}
  {\bibfnamefont {Q.}~\bibnamefont {Liu}}, \bibinfo {author} {\bibfnamefont
  {Y.}~\bibnamefont {Zhao}}, \bibinfo {author} {\bibfnamefont {H.-Z.}\
  \bibnamefont {Lu}}, \ and\ \bibinfo {author} {\bibfnamefont {X.~C.}\
  \bibnamefont {Xie}},\ }\bibfield  {title} {\enquote {\bibinfo {title} {Using
  nonlocal surface transport to identify the axion insulator}}, }\href
  {\doibase 10.1103/PhysRevB.103.L241409} {\bibfield  {journal} {\bibinfo
  {journal} {Phys. Rev. B}\ }\textbf {\bibinfo {volume} {103}},\ \bibinfo
  {pages} {L241409} (\bibinfo {year} {2021}{\natexlab{a}})}\BibitemShut
  {NoStop}%
\bibitem [{\citenamefont {Zhang}\ \emph {et~al.}(2014)\citenamefont {Zhang},
  \citenamefont {Liu}, \citenamefont {Luo}, \citenamefont {Freeman},\ and\
  \citenamefont {Zunger}}]{Zhang2014NatPhys}%
  \BibitemOpen
  \bibfield  {author} {\bibinfo {author} {\bibfnamefont {X.}~\bibnamefont
  {Zhang}}, \bibinfo {author} {\bibfnamefont {Q.}~\bibnamefont {Liu}}, \bibinfo
  {author} {\bibfnamefont {J.-W.}\ \bibnamefont {Luo}}, \bibinfo {author}
  {\bibfnamefont {A.~J.}\ \bibnamefont {Freeman}}, \ and\ \bibinfo {author}
  {\bibfnamefont {A.}~\bibnamefont {Zunger}},\ }\bibfield  {title} {\enquote
  {\bibinfo {title} {Hidden spin polarization in inversion-symmetric bulk
  crystals}}, }\href {\doibase 10.1038/nphys2933} {\bibfield  {journal}
  {\bibinfo  {journal} {Nat. Phys.}\ }\textbf {\bibinfo {volume} {10}},\
  \bibinfo {pages} {387} (\bibinfo {year} {2014})}\BibitemShut {NoStop}%
\bibitem [{\citenamefont {Beaulieu}\ \emph {et~al.}(2020)\citenamefont
  {Beaulieu}, \citenamefont {Schusser}, \citenamefont {Dong}, \citenamefont
  {Sch\"uler}, \citenamefont {Pincelli}, \citenamefont {Dendzik}, \citenamefont
  {Maklar}, \citenamefont {Neef}, \citenamefont {Ebert}, \citenamefont
  {Hricovini}, \citenamefont {Wolf}, \citenamefont {Braun}, \citenamefont
  {Rettig}, \citenamefont {Min\'ar},\ and\ \citenamefont
  {Ernstorfer}}]{Beaulieu2020PRL}%
  \BibitemOpen
  \bibfield  {author} {\bibinfo {author} {\bibfnamefont {S.}~\bibnamefont
  {Beaulieu}}, \bibinfo {author} {\bibfnamefont {J.}~\bibnamefont {Schusser}},
  \bibinfo {author} {\bibfnamefont {S.}~\bibnamefont {Dong}}, \bibinfo {author}
  {\bibfnamefont {M.}~\bibnamefont {Sch\"uler}}, \bibinfo {author}
  {\bibfnamefont {T.}~\bibnamefont {Pincelli}}, \bibinfo {author}
  {\bibfnamefont {M.}~\bibnamefont {Dendzik}},  \emph {et~al.},\ }\bibfield
  {title} {\enquote {\bibinfo {title} {Revealing hidden orbital pseudospin
  texture with time-reversal dichroism in photoelectron angular
  distributions}}, }\href {\doibase 10.1103/PhysRevLett.125.216404} {\bibfield
  {journal} {\bibinfo  {journal} {Phys. Rev. Lett.}\ }\textbf {\bibinfo
  {volume} {125}},\ \bibinfo {pages} {216404} (\bibinfo {year}
  {2020})}\BibitemShut {NoStop}%
\bibitem [{\citenamefont {Razzoli}\ \emph {et~al.}(2017)\citenamefont
  {Razzoli}, \citenamefont {Jaouen}, \citenamefont {Mottas}, \citenamefont
  {Hildebrand}, \citenamefont {Monney}, \citenamefont {Pisoni}, \citenamefont
  {Muff}, \citenamefont {Fanciulli}, \citenamefont {Plumb}, \citenamefont
  {Rogalev}, \citenamefont {Strocov}, \citenamefont {Mesot}, \citenamefont
  {Shi}, \citenamefont {Dil}, \citenamefont {Beck},\ and\ \citenamefont
  {Aebi}}]{Razzoli2017PRL}%
  \BibitemOpen
  \bibfield  {author} {\bibinfo {author} {\bibfnamefont {E.}~\bibnamefont
  {Razzoli}}, \bibinfo {author} {\bibfnamefont {T.}~\bibnamefont {Jaouen}},
  \bibinfo {author} {\bibfnamefont {M.-L.}\ \bibnamefont {Mottas}}, \bibinfo
  {author} {\bibfnamefont {B.}~\bibnamefont {Hildebrand}}, \bibinfo {author}
  {\bibfnamefont {G.}~\bibnamefont {Monney}}, \bibinfo {author} {\bibfnamefont
  {A.}~\bibnamefont {Pisoni}},  \emph {et~al.},\ }\bibfield  {title} {\enquote
  {\bibinfo {title} {{Selective probing of hidden spin-polarized states in
  inversion-symmetric bulk MoS$_2$}}}, }\href {\doibase
  10.1103/PhysRevLett.118.086402} {\bibfield  {journal} {\bibinfo  {journal}
  {Phys. Rev. Lett.}\ }\textbf {\bibinfo {volume} {118}},\ \bibinfo {pages}
  {086402} (\bibinfo {year} {2017})}\BibitemShut {NoStop}%
\bibitem [{\citenamefont {Cho}\ \emph {et~al.}(2018)\citenamefont {Cho},
  \citenamefont {Park}, \citenamefont {Hong}, \citenamefont {Jung},
  \citenamefont {Kim}, \citenamefont {Han}, \citenamefont {Kyung},
  \citenamefont {Kim}, \citenamefont {Mo}, \citenamefont {Denlinger},
  \citenamefont {Shim}, \citenamefont {Han}, \citenamefont {Kim},\ and\
  \citenamefont {Park}}]{Cho2018PRL}%
  \BibitemOpen
  \bibfield  {author} {\bibinfo {author} {\bibfnamefont {S.}~\bibnamefont
  {Cho}}, \bibinfo {author} {\bibfnamefont {J.-H.}\ \bibnamefont {Park}},
  \bibinfo {author} {\bibfnamefont {J.}~\bibnamefont {Hong}}, \bibinfo {author}
  {\bibfnamefont {J.}~\bibnamefont {Jung}}, \bibinfo {author} {\bibfnamefont
  {B.~S.}\ \bibnamefont {Kim}}, \bibinfo {author} {\bibfnamefont
  {G.}~\bibnamefont {Han}},  \emph {et~al.},\ }\bibfield  {title} {\enquote
  {\bibinfo {title} {{Experimental observation of hidden Berry curvature in
  inversion-symmetric bulk $2H\text{\ensuremath{-}}{\mathrm{WSe}}_{2}$}}},
  }\href {\doibase 10.1103/PhysRevLett.121.186401} {\bibfield  {journal}
  {\bibinfo  {journal} {Phys. Rev. Lett.}\ }\textbf {\bibinfo {volume} {121}},\
  \bibinfo {pages} {186401} (\bibinfo {year} {2018})}\BibitemShut {NoStop}%
\bibitem [{\citenamefont {Schüler}\ \emph {et~al.}(2020)\citenamefont
  {Schüler}, \citenamefont {Giovannini}, \citenamefont {Hübener},
  \citenamefont {Rubio}, \citenamefont {Sentef},\ and\ \citenamefont
  {Werner}}]{Schuler2020SciAdv}%
  \BibitemOpen
  \bibfield  {author} {\bibinfo {author} {\bibfnamefont {M.}~\bibnamefont
  {Schüler}}, \bibinfo {author} {\bibfnamefont {U.~D.}\ \bibnamefont
  {Giovannini}}, \bibinfo {author} {\bibfnamefont {H.}~\bibnamefont
  {Hübener}}, \bibinfo {author} {\bibfnamefont {A.}~\bibnamefont {Rubio}},
  \bibinfo {author} {\bibfnamefont {M.~A.}\ \bibnamefont {Sentef}}, \ and\
  \bibinfo {author} {\bibfnamefont {P.}~\bibnamefont {Werner}},\ }\bibfield
  {title} {\enquote {\bibinfo {title} {{Local Berry curvature signatures in
  dichroic angle-resolved photoelectron spectroscopy from two-dimensional
  materials}}}, }\href {\doibase 10.1126/sciadv.aay2730} {\bibfield  {journal}
  {\bibinfo  {journal} {Sci. Adv.}\ }\textbf {\bibinfo {volume} {6}},\ \bibinfo
  {pages} {eaay2730} (\bibinfo {year} {2020})}\BibitemShut {NoStop}%
\bibitem [{\citenamefont {Murakawa}\ \emph {et~al.}(2013)\citenamefont
  {Murakawa}, \citenamefont {Bahramy}, \citenamefont {Tokunaga}, \citenamefont
  {Kohama}, \citenamefont {Bell}, \citenamefont {Kaneko}, \citenamefont
  {Nagaosa}, \citenamefont {Hwang},\ and\ \citenamefont
  {Tokura}}]{Murakawa2013}%
  \BibitemOpen
  \bibfield  {author} {\bibinfo {author} {\bibfnamefont {H.}~\bibnamefont
  {Murakawa}}, \bibinfo {author} {\bibfnamefont {M.~S.}\ \bibnamefont
  {Bahramy}}, \bibinfo {author} {\bibfnamefont {M.}~\bibnamefont {Tokunaga}},
  \bibinfo {author} {\bibfnamefont {Y.}~\bibnamefont {Kohama}}, \bibinfo
  {author} {\bibfnamefont {C.}~\bibnamefont {Bell}}, \bibinfo {author}
  {\bibfnamefont {Y.}~\bibnamefont {Kaneko}}, \bibinfo {author} {\bibfnamefont
  {N.}~\bibnamefont {Nagaosa}}, \bibinfo {author} {\bibfnamefont {H.~Y.}\
  \bibnamefont {Hwang}}, \ and\ \bibinfo {author} {\bibfnamefont
  {Y.}~\bibnamefont {Tokura}},\ }\bibfield  {title} {\enquote {\bibinfo {title}
  {Detection of {Berry's} phase in a bulk {Rashba} semiconductor}}, }\href
  {\doibase 10.1126/science.1242247} {\bibfield  {journal} {\bibinfo  {journal}
  {Science}\ }\textbf {\bibinfo {volume} {342}},\ \bibinfo {pages} {1490}
  (\bibinfo {year} {2013})}\BibitemShut {NoStop}%
\bibitem [{\citenamefont {Zhang}\ \emph {et~al.}(2019)\citenamefont {Zhang},
  \citenamefont {Shi}, \citenamefont {Zhu}, \citenamefont {Xing}, \citenamefont
  {Zhang},\ and\ \citenamefont {Wang}}]{Zhang19prl}%
  \BibitemOpen
  \bibfield  {author} {\bibinfo {author} {\bibfnamefont {D.}~\bibnamefont
  {Zhang}}, \bibinfo {author} {\bibfnamefont {M.}~\bibnamefont {Shi}}, \bibinfo
  {author} {\bibfnamefont {T.}~\bibnamefont {Zhu}}, \bibinfo {author}
  {\bibfnamefont {D.}~\bibnamefont {Xing}}, \bibinfo {author} {\bibfnamefont
  {H.}~\bibnamefont {Zhang}}, \ and\ \bibinfo {author} {\bibfnamefont
  {J.}~\bibnamefont {Wang}},\ }\bibfield  {title} {\enquote {\bibinfo {title}
  {Topological axion states in the magnetic insulator {MnBi$_{2}$Te$_{4}$} with
  the quantized magnetoelectric effect}}, }\href
  {https://link.aps.org/doi/10.1103/PhysRevLett.122.206401} {\bibfield
  {journal} {\bibinfo  {journal} {Phys. Rev. Lett.}\ }\textbf {\bibinfo
  {volume} {122}},\ \bibinfo {pages} {206401} (\bibinfo {year}
  {2019})}\BibitemShut {NoStop}%
\bibitem [{\citenamefont {Zhang}\ \emph
  {et~al.}(2020{\natexlab{a}})\citenamefont {Zhang}, \citenamefont {Wu},\ and\
  \citenamefont {{Das Sarma}}}]{Zhang20prl}%
  \BibitemOpen
  \bibfield  {author} {\bibinfo {author} {\bibfnamefont {R.-X.}\ \bibnamefont
  {Zhang}}, \bibinfo {author} {\bibfnamefont {F.}~\bibnamefont {Wu}}, \ and\
  \bibinfo {author} {\bibfnamefont {S.}~\bibnamefont {{Das Sarma}}},\
  }\bibfield  {title} {\enquote {\bibinfo {title} {M\"obius insulator and
  higher-order topology in {MnBi}$_{2n}${Te}$_{3n+1}$}}, }\href {\doibase
  10.1103/physrevlett.124.136407} {\bibfield  {journal} {\bibinfo  {journal}
  {Phys. Rev. Lett.}\ }\textbf {\bibinfo {volume} {124}},\ \bibinfo {pages}
  {136407} (\bibinfo {year} {2020}{\natexlab{a}})}\BibitemShut {NoStop}%
\bibitem [{\citenamefont {Li}\ \emph {et~al.}(2021)\citenamefont {Li},
  \citenamefont {Jiang}, \citenamefont {Chen},\ and\ \citenamefont
  {Xie}}]{LiHailong21PRL}%
  \BibitemOpen
  \bibfield  {author} {\bibinfo {author} {\bibfnamefont {H.}~\bibnamefont
  {Li}}, \bibinfo {author} {\bibfnamefont {H.}~\bibnamefont {Jiang}}, \bibinfo
  {author} {\bibfnamefont {C.-Z.}\ \bibnamefont {Chen}}, \ and\ \bibinfo
  {author} {\bibfnamefont {X.~C.}\ \bibnamefont {Xie}},\ }\bibfield  {title}
  {\enquote {\bibinfo {title} {Critical behavior and universal signature of an
  axion insulator state}}, }\href {\doibase 10.1103/PhysRevLett.126.156601}
  {\bibfield  {journal} {\bibinfo  {journal} {Phys. Rev. Lett.}\ }\textbf
  {\bibinfo {volume} {126}},\ \bibinfo {pages} {156601} (\bibinfo {year}
  {2021})}\BibitemShut {NoStop}%
\bibitem [{\citenamefont {Ding}\ \emph {et~al.}(2020)\citenamefont {Ding},
  \citenamefont {Xu}, \citenamefont {Chen},\ and\ \citenamefont
  {Xie}}]{Ding20prbrc}%
  \BibitemOpen
  \bibfield  {author} {\bibinfo {author} {\bibfnamefont {Y.-R.}\ \bibnamefont
  {Ding}}, \bibinfo {author} {\bibfnamefont {D.-H.}\ \bibnamefont {Xu}},
  \bibinfo {author} {\bibfnamefont {C.-Z.}\ \bibnamefont {Chen}}, \ and\
  \bibinfo {author} {\bibfnamefont {X.~C.}\ \bibnamefont {Xie}},\ }\bibfield
  {title} {\enquote {\bibinfo {title} {{Hinged quantum spin Hall effect in
  antiferromagnetic topological insulators}}}, }\href {\doibase
  10.1103/physrevb.101.041404} {\bibfield  {journal} {\bibinfo  {journal}
  {Phys. Rev. B}\ }\textbf {\bibinfo {volume} {101}},\ \bibinfo {pages}
  {041404(R)} (\bibinfo {year} {2020})}\BibitemShut {NoStop}%
\bibitem [{\citenamefont {Chen}\ \emph
  {et~al.}(2021{\natexlab{b}})\citenamefont {Chen}, \citenamefont {Qi},
  \citenamefont {Xu},\ and\ \citenamefont {Xie}}]{ChenCZ2021arXiv}%
  \BibitemOpen
  \bibfield  {author} {\bibinfo {author} {\bibfnamefont {C.-Z.}\ \bibnamefont
  {Chen}}, \bibinfo {author} {\bibfnamefont {J.}~\bibnamefont {Qi}}, \bibinfo
  {author} {\bibfnamefont {D.-H.}\ \bibnamefont {Xu}}, \ and\ \bibinfo {author}
  {\bibfnamefont {X.}~\bibnamefont {Xie}},\ }\bibfield  {title} {\enquote
  {\bibinfo {title} {{Evolution of Berry curvature and reentrant quantum
  anomalous Hall effect in an intrinsic magnetic topological insulator}}},
  }\href {\doibase 10.1007/s11433-021-1774-1} {\bibfield  {journal} {\bibinfo
  {journal} {Sci. China Phys. Mech. Astron.}\ }\textbf {\bibinfo {volume}
  {64}},\ \bibinfo {pages} {127211} (\bibinfo {year}
  {2021}{\natexlab{b}})}\BibitemShut {NoStop}%
\bibitem [{\citenamefont {Prodan}(2011)}]{Prodan2011JPA}%
  \BibitemOpen
  \bibfield  {author} {\bibinfo {author} {\bibfnamefont {E.}~\bibnamefont
  {Prodan}},\ }\bibfield  {title} {\enquote {\bibinfo {title} {Disordered
  topological insulators: a non-commutative geometry perspective}}, }\href
  {\doibase 10.1088/1751-8113/44/11/113001} {\bibfield  {journal} {\bibinfo
  {journal} {J. Phys. A: Math. Theor.}\ }\textbf {\bibinfo {volume} {44}},\
  \bibinfo {pages} {113001} (\bibinfo {year} {2011})}\BibitemShut {NoStop}%
\bibitem [{\citenamefont {Bianco}\ and\ \citenamefont
  {Resta}(2011)}]{PhysRevB.84.241106}%
  \BibitemOpen
  \bibfield  {author} {\bibinfo {author} {\bibfnamefont {R.}~\bibnamefont
  {Bianco}}\ and\ \bibinfo {author} {\bibfnamefont {R.}~\bibnamefont {Resta}},\
  }\bibfield  {title} {\enquote {\bibinfo {title} {Mapping topological order in
  coordinate space}}, }\href {\doibase 10.1103/PhysRevB.84.241106} {\bibfield
  {journal} {\bibinfo  {journal} {Phys. Rev. B}\ }\textbf {\bibinfo {volume}
  {84}},\ \bibinfo {pages} {241106(R)} (\bibinfo {year} {2011})}\BibitemShut
  {NoStop}%
\bibitem [{\citenamefont {Caio}\ \emph {et~al.}(2019)\citenamefont {Caio},
  \citenamefont {Moller}, \citenamefont {Cooper},\ and\ \citenamefont
  {Bhaseen}}]{Caio2019}%
  \BibitemOpen
  \bibfield  {author} {\bibinfo {author} {\bibfnamefont {M.~D.}\ \bibnamefont
  {Caio}}, \bibinfo {author} {\bibfnamefont {G.}~\bibnamefont {Moller}},
  \bibinfo {author} {\bibfnamefont {N.~R.}\ \bibnamefont {Cooper}}, \ and\
  \bibinfo {author} {\bibfnamefont {M.~J.}\ \bibnamefont {Bhaseen}},\
  }\bibfield  {title} {\enquote {\bibinfo {title} {{Topological marker currents
  in Chern insulators}}}, }\href {\doibase 10.1038/s41567-018-0390-7}
  {\bibfield  {journal} {\bibinfo  {journal} {Nat. Phys.}\ }\textbf {\bibinfo
  {volume} {15}},\ \bibinfo {pages} {257} (\bibinfo {year} {2019})}\BibitemShut
  {NoStop}%
\bibitem [{\citenamefont {Varnava}\ and\ \citenamefont
  {Vanderbilt}(2018)}]{Varnava18prb}%
  \BibitemOpen
  \bibfield  {author} {\bibinfo {author} {\bibfnamefont {N.}~\bibnamefont
  {Varnava}}\ and\ \bibinfo {author} {\bibfnamefont {D.}~\bibnamefont
  {Vanderbilt}},\ }\bibfield  {title} {\enquote {\bibinfo {title} {{Surfaces of
  axion insulators}}}, }\href {\doibase 10.1103/physrevb.98.245117} {\bibfield
  {journal} {\bibinfo  {journal} {Phys. Rev. B}\ }\textbf {\bibinfo {volume}
  {98}},\ \bibinfo {pages} {245117} (\bibinfo {year} {2018})}\BibitemShut
  {NoStop}%
\bibitem [{\citenamefont {Zhang}\ \emph
  {et~al.}(2020{\natexlab{b}})\citenamefont {Zhang}, \citenamefont {Wang},
  \citenamefont {Shi}, \citenamefont {Zhu}, \citenamefont {Zhang},\ and\
  \citenamefont {Wang}}]{JZhang2020CPL}%
  \BibitemOpen
  \bibfield  {author} {\bibinfo {author} {\bibfnamefont {J.}~\bibnamefont
  {Zhang}}, \bibinfo {author} {\bibfnamefont {D.}~\bibnamefont {Wang}},
  \bibinfo {author} {\bibfnamefont {M.}~\bibnamefont {Shi}}, \bibinfo {author}
  {\bibfnamefont {T.}~\bibnamefont {Zhu}}, \bibinfo {author} {\bibfnamefont
  {H.}~\bibnamefont {Zhang}}, \ and\ \bibinfo {author} {\bibfnamefont
  {J.}~\bibnamefont {Wang}},\ }\bibfield  {title} {\enquote {\bibinfo {title}
  {Large dynamical axion field in topological antiferromagnetic insulator
  {Mn$_2$Bi$_2$Te$_5$}}}, }\href {\doibase 10.1088/0256-307x/37/7/077304}
  {\bibfield  {journal} {\bibinfo  {journal} {Chin. Phys. Lett.}\ }\textbf
  {\bibinfo {volume} {37}},\ \bibinfo {pages} {077304} (\bibinfo {year}
  {2020}{\natexlab{b}})}\BibitemShut {NoStop}%
\bibitem [{\citenamefont {Wang}\ \emph {et~al.}(2015)\citenamefont {Wang},
  \citenamefont {Lian}, \citenamefont {Qi},\ and\ \citenamefont
  {Zhang}}]{Wang15prbrc}%
  \BibitemOpen
  \bibfield  {author} {\bibinfo {author} {\bibfnamefont {J.}~\bibnamefont
  {Wang}}, \bibinfo {author} {\bibfnamefont {B.}~\bibnamefont {Lian}}, \bibinfo
  {author} {\bibfnamefont {X.-L.}\ \bibnamefont {Qi}}, \ and\ \bibinfo {author}
  {\bibfnamefont {S.-C.}\ \bibnamefont {Zhang}},\ }\bibfield  {title} {\enquote
  {\bibinfo {title} {Quantized topological magnetoelectric effect of the
  zero-plateau quantum anomalous {Hall} state}}, }\href
  {https://link.aps.org/doi/10.1103/PhysRevB.92.081107} {\bibfield  {journal}
  {\bibinfo  {journal} {Phys. Rev. B}\ }\textbf {\bibinfo {volume} {92}},\
  \bibinfo {pages} {081107(R)} (\bibinfo {year} {2015})}\BibitemShut {NoStop}%
\bibitem [{\citenamefont {Liu}\ \emph {et~al.}(2020)\citenamefont {Liu},
  \citenamefont {Wang}, \citenamefont {Li}, \citenamefont {Wu}, \citenamefont
  {Li}, \citenamefont {Li}, \citenamefont {He}, \citenamefont {Xu},
  \citenamefont {Zhang},\ and\ \citenamefont {Wang}}]{Liu20nm}%
  \BibitemOpen
  \bibfield  {author} {\bibinfo {author} {\bibfnamefont {C.}~\bibnamefont
  {Liu}}, \bibinfo {author} {\bibfnamefont {Y.}~\bibnamefont {Wang}}, \bibinfo
  {author} {\bibfnamefont {H.}~\bibnamefont {Li}}, \bibinfo {author}
  {\bibfnamefont {Y.}~\bibnamefont {Wu}}, \bibinfo {author} {\bibfnamefont
  {Y.}~\bibnamefont {Li}}, \bibinfo {author} {\bibfnamefont {J.}~\bibnamefont
  {Li}}, \bibinfo {author} {\bibfnamefont {K.}~\bibnamefont {He}}, \bibinfo
  {author} {\bibfnamefont {Y.}~\bibnamefont {Xu}}, \bibinfo {author}
  {\bibfnamefont {J.}~\bibnamefont {Zhang}}, \ and\ \bibinfo {author}
  {\bibfnamefont {Y.}~\bibnamefont {Wang}},\ }\bibfield  {title} {\enquote
  {\bibinfo {title} {Robust axion insulator and {Chern} insulator phases in a
  two-dimensional antiferromagnetic topological insulator}}, }\href
  {https://doi.org/10.1038/s41563-019-0573-3} {\bibfield  {journal} {\bibinfo
  {journal} {Nat. Mater.}\ }\textbf {\bibinfo {volume} {19}},\ \bibinfo {pages}
  {522} (\bibinfo {year} {2020})}\BibitemShut {NoStop}%
\bibitem [{\citenamefont {Mogi}\ \emph {et~al.}(2017)\citenamefont {Mogi},
  \citenamefont {Kawamura}, \citenamefont {Yoshimi}, \citenamefont {Tsukazaki},
  \citenamefont {Kozuka}, \citenamefont {Shirakawa}, \citenamefont {Takahashi},
  \citenamefont {Kawasaki},\ and\ \citenamefont {Tokura}}]{Mogi2017NatMat}%
  \BibitemOpen
  \bibfield  {author} {\bibinfo {author} {\bibfnamefont {M.}~\bibnamefont
  {Mogi}}, \bibinfo {author} {\bibfnamefont {M.}~\bibnamefont {Kawamura}},
  \bibinfo {author} {\bibfnamefont {R.}~\bibnamefont {Yoshimi}}, \bibinfo
  {author} {\bibfnamefont {A.}~\bibnamefont {Tsukazaki}}, \bibinfo {author}
  {\bibfnamefont {Y.}~\bibnamefont {Kozuka}}, \bibinfo {author} {\bibfnamefont
  {N.}~\bibnamefont {Shirakawa}}, \bibinfo {author} {\bibfnamefont {K.~S.}\
  \bibnamefont {Takahashi}}, \bibinfo {author} {\bibfnamefont {M.}~\bibnamefont
  {Kawasaki}}, \ and\ \bibinfo {author} {\bibfnamefont {Y.}~\bibnamefont
  {Tokura}},\ }\bibfield  {title} {\enquote {\bibinfo {title} {A magnetic
  heterostructure of topological insulators as a candidate for an axion
  insulator}}, }\href {\doibase 10.1038/nmat4855} {\bibfield  {journal}
  {\bibinfo  {journal} {Nat. Mater.}\ }\textbf {\bibinfo {volume} {16}},\
  \bibinfo {pages} {516} (\bibinfo {year} {2017})}\BibitemShut {NoStop}%
\bibitem [{\citenamefont {Gu}\ \emph {et~al.}(2021)\citenamefont {Gu},
  \citenamefont {Li}, \citenamefont {Sun}, \citenamefont {Zhao}, \citenamefont
  {Liu}, \citenamefont {Liu}, \citenamefont {Lu},\ and\ \citenamefont
  {Liu}}]{Gu2020Arxiv}%
  \BibitemOpen
  \bibfield  {author} {\bibinfo {author} {\bibfnamefont {M.}~\bibnamefont
  {Gu}}, \bibinfo {author} {\bibfnamefont {J.}~\bibnamefont {Li}}, \bibinfo
  {author} {\bibfnamefont {H.}~\bibnamefont {Sun}}, \bibinfo {author}
  {\bibfnamefont {Y.}~\bibnamefont {Zhao}}, \bibinfo {author} {\bibfnamefont
  {C.}~\bibnamefont {Liu}}, \bibinfo {author} {\bibfnamefont {J.}~\bibnamefont
  {Liu}}, \bibinfo {author} {\bibfnamefont {H.}~\bibnamefont {Lu}}, \ and\
  \bibinfo {author} {\bibfnamefont {Q.}~\bibnamefont {Liu}},\ }\bibfield
  {title} {\enquote {\bibinfo {title} {{Spectral signatures of the surface
  anomalous Hall effect in magnetic axion insulators}}}, }\href {\doibase
  10.1038/s41467-021-23844-z} {\bibfield  {journal} {\bibinfo  {journal} {Nat.
  Commun.}\ }\textbf {\bibinfo {volume} {12}},\ \bibinfo {pages} {3524}
  (\bibinfo {year} {2021})}\BibitemShut {NoStop}%
\bibitem [{\citenamefont {Kresse}\ and\ \citenamefont
  {Furthmuller}(1996)}]{Kresse/Furthmuller:1996a}%
  \BibitemOpen
  \bibfield  {author} {\bibinfo {author} {\bibfnamefont {G.}~\bibnamefont
  {Kresse}}\ and\ \bibinfo {author} {\bibfnamefont {J.}~\bibnamefont
  {Furthmuller}},\ }\bibfield  {title} {\enquote {\bibinfo {title} {Efficiency
  of ab-initio total energy calculations for metals and semiconductors using a
  plane-wave basis set}}, }\href {\doibase 10.1016/0927-0256(96)00008-0}
  {\bibfield  {journal} {\bibinfo  {journal} {Comput. Phys. Commun.}\ }\textbf
  {\bibinfo {volume} {6}},\ \bibinfo {pages} {15} (\bibinfo {year}
  {1996})}\BibitemShut {NoStop}%
\bibitem [{\citenamefont {Kresse}\ and\ \citenamefont
  {Joubert}(1999)}]{Kresse/Joubert:1999}%
  \BibitemOpen
  \bibfield  {author} {\bibinfo {author} {\bibfnamefont {G.}~\bibnamefont
  {Kresse}}\ and\ \bibinfo {author} {\bibfnamefont {D.}~\bibnamefont
  {Joubert}},\ }\bibfield  {title} {\enquote {\bibinfo {title} {From ultrasoft
  pseudopotentials to the projector augmented-wave method}}, }\href {\doibase
  10.1103/PhysRevB.59.1758} {\bibfield  {journal} {\bibinfo  {journal} {Phys.
  Rev. B}\ }\textbf {\bibinfo {volume} {59}},\ \bibinfo {pages} {1758}
  (\bibinfo {year} {1999})}\BibitemShut {NoStop}%
\bibitem [{\citenamefont {Bl\"ochl}(1994)}]{Blochl:1994}%
  \BibitemOpen
  \bibfield  {author} {\bibinfo {author} {\bibfnamefont {P.~E.}\ \bibnamefont
  {Bl\"ochl}},\ }\bibfield  {title} {\enquote {\bibinfo {title} {Projector
  augmented-wave method}}, }\href {\doibase 10.1103/PhysRevB.50.17953}
  {\bibfield  {journal} {\bibinfo  {journal} {Phys. Rev. B}\ }\textbf {\bibinfo
  {volume} {50}},\ \bibinfo {pages} {17953} (\bibinfo {year}
  {1994})}\BibitemShut {NoStop}%
\bibitem [{\citenamefont {Perdew}\ \emph {et~al.}(1996)\citenamefont {Perdew},
  \citenamefont {Burke},\ and\ \citenamefont {Ernzerhof}}]{PBE}%
  \BibitemOpen
  \bibfield  {author} {\bibinfo {author} {\bibfnamefont {J.~P.}\ \bibnamefont
  {Perdew}}, \bibinfo {author} {\bibfnamefont {K.}~\bibnamefont {Burke}}, \
  and\ \bibinfo {author} {\bibfnamefont {M.}~\bibnamefont {Ernzerhof}},\
  }\bibfield  {title} {\enquote {\bibinfo {title} {Generalized gradient
  approximation made simple}}, }\href {\doibase 10.1103/physrevlett.77.3865}
  {\bibfield  {journal} {\bibinfo  {journal} {Phys. Rev. Lett.}\ }\textbf
  {\bibinfo {volume} {77}},\ \bibinfo {pages} {3865} (\bibinfo {year}
  {1996})}\BibitemShut {NoStop}%
\bibitem [{\citenamefont {Dudarev}\ \emph {et~al.}(2000)\citenamefont
  {Dudarev}, \citenamefont {Peng}, \citenamefont {Savrasov},\ and\
  \citenamefont {Zuo}}]{Dudarev/Zuo_et_al:2506}%
  \BibitemOpen
  \bibfield  {author} {\bibinfo {author} {\bibfnamefont {S.~L.}\ \bibnamefont
  {Dudarev}}, \bibinfo {author} {\bibfnamefont {L.-M.}\ \bibnamefont {Peng}},
  \bibinfo {author} {\bibfnamefont {S.~Y.}\ \bibnamefont {Savrasov}}, \ and\
  \bibinfo {author} {\bibfnamefont {J.-M.}\ \bibnamefont {Zuo}},\ }\bibfield
  {title} {\enquote {\bibinfo {title} {{Correlation effects in the ground-state
  charge density of Mott insulating NiO: A comparison of ab initio calculations
  and high-energy electron diffraction measurements}}}, }\href {\doibase
  10.1103/PhysRevB.61.2506} {\bibfield  {journal} {\bibinfo  {journal} {Phys.
  Rev. B}\ }\textbf {\bibinfo {volume} {61}},\ \bibinfo {pages} {2506}
  (\bibinfo {year} {2000})}\BibitemShut {NoStop}%
\bibitem [{\citenamefont {Sun}\ \emph {et~al.}(2020)\citenamefont {Sun},
  \citenamefont {Wang}, \citenamefont {Zhang}, \citenamefont {Chen},
  \citenamefont {Zhao}, \citenamefont {Liu}, \citenamefont {Liu}, \citenamefont
  {Chen}, \citenamefont {Lu},\ and\ \citenamefont {Xie}}]{Sun20prbrc}%
  \BibitemOpen
  \bibfield  {author} {\bibinfo {author} {\bibfnamefont {H.-P.}\ \bibnamefont
  {Sun}}, \bibinfo {author} {\bibfnamefont {C.~M.}\ \bibnamefont {Wang}},
  \bibinfo {author} {\bibfnamefont {S.-B.}\ \bibnamefont {Zhang}}, \bibinfo
  {author} {\bibfnamefont {R.}~\bibnamefont {Chen}}, \bibinfo {author}
  {\bibfnamefont {Y.}~\bibnamefont {Zhao}}, \bibinfo {author} {\bibfnamefont
  {C.}~\bibnamefont {Liu}}, \bibinfo {author} {\bibfnamefont {Q.}~\bibnamefont
  {Liu}}, \bibinfo {author} {\bibfnamefont {C.}~\bibnamefont {Chen}}, \bibinfo
  {author} {\bibfnamefont {H.-Z.}\ \bibnamefont {Lu}}, \ and\ \bibinfo {author}
  {\bibfnamefont {X.~C.}\ \bibnamefont {Xie}},\ }\bibfield  {title} {\enquote
  {\bibinfo {title} {{Analytical solution for the surface states of the
  antiferromagnetic topological insulator MnBi$_2$Te$_4$}}}, }\href {\doibase
  10.1103/PhysRevB.102.241406} {\bibfield  {journal} {\bibinfo  {journal}
  {Phys. Rev. B}\ }\textbf {\bibinfo {volume} {102}},\ \bibinfo {pages}
  {241406(R)} (\bibinfo {year} {2020})}\BibitemShut {NoStop}%
\bibitem [{\citenamefont {Shikin}\ \emph {et~al.}(2020)\citenamefont {Shikin},
  \citenamefont {Estyunin}, \citenamefont {Klimovskikh}, \citenamefont
  {Filnov}, \citenamefont {Schwier}, \citenamefont {Kumar}, \citenamefont
  {Miyamoto}, \citenamefont {Okuda}, \citenamefont {Kimura}, \citenamefont
  {Kuroda}, \citenamefont {Yaji}, \citenamefont {Shin}, \citenamefont {Takeda},
  \citenamefont {Saitoh}, \citenamefont {Aliev}, \citenamefont {Mamedov},
  \citenamefont {Amiraslanov}, \citenamefont {Babanly}, \citenamefont
  {Otrokov}, \citenamefont {Eremeev},\ and\ \citenamefont
  {Chulkov}}]{Shikin2020scirep}%
  \BibitemOpen
  \bibfield  {author} {\bibinfo {author} {\bibfnamefont {A.~M.}\ \bibnamefont
  {Shikin}}, \bibinfo {author} {\bibfnamefont {D.~A.}\ \bibnamefont
  {Estyunin}}, \bibinfo {author} {\bibfnamefont {I.~I.}\ \bibnamefont
  {Klimovskikh}}, \bibinfo {author} {\bibfnamefont {S.~O.}\ \bibnamefont
  {Filnov}}, \bibinfo {author} {\bibfnamefont {E.~F.}\ \bibnamefont {Schwier}},
  \bibinfo {author} {\bibfnamefont {S.}~\bibnamefont {Kumar}},  \emph
  {et~al.},\ }\bibfield  {title} {\enquote {\bibinfo {title} {{Nature of the
  Dirac gap modulation and surface magnetic interaction in axion
  antiferromagnetic topological insulator MnBi$_2$Te$_4$}}}, }\href {\doibase
  10.1038/s41598-020-70089-9} {\bibfield  {journal} {\bibinfo  {journal} {Sci.
  Rep.}\ }\textbf {\bibinfo {volume} {10}},\ \bibinfo {pages} {13226} (\bibinfo
  {year} {2020})}\BibitemShut {NoStop}%
\bibitem [{\citenamefont {Mostofi}\ \emph {et~al.}(2008)\citenamefont
  {Mostofi}, \citenamefont {Yates}, \citenamefont {Lee}, \citenamefont {Souza},
  \citenamefont {Vanderbilt},\ and\ \citenamefont {Marzari}}]{wannier90}%
  \BibitemOpen
  \bibfield  {author} {\bibinfo {author} {\bibfnamefont {A.~A.}\ \bibnamefont
  {Mostofi}}, \bibinfo {author} {\bibfnamefont {J.~R.}\ \bibnamefont {Yates}},
  \bibinfo {author} {\bibfnamefont {Y.-S.}\ \bibnamefont {Lee}}, \bibinfo
  {author} {\bibfnamefont {I.}~\bibnamefont {Souza}}, \bibinfo {author}
  {\bibfnamefont {D.}~\bibnamefont {Vanderbilt}}, \ and\ \bibinfo {author}
  {\bibfnamefont {N.}~\bibnamefont {Marzari}},\ }\bibfield  {title} {\enquote
  {\bibinfo {title} {{Wannier90: A tool for obtaining maximally-localised
  Wannier functions}}}, }\href {\doibase 10.1016/j.cpc.2007.11.016} {\bibfield
  {journal} {\bibinfo  {journal} {Comput. Phys. Commun.}\ }\textbf {\bibinfo
  {volume} {178}},\ \bibinfo {pages} {685} (\bibinfo {year}
  {2008})}\BibitemShut {NoStop}%
\bibitem [{\citenamefont {Wu}\ \emph {et~al.}(2018)\citenamefont {Wu},
  \citenamefont {Zhang}, \citenamefont {Song}, \citenamefont {Troyer},\ and\
  \citenamefont {Soluyanov}}]{wanniertools}%
  \BibitemOpen
  \bibfield  {author} {\bibinfo {author} {\bibfnamefont {Q.}~\bibnamefont
  {Wu}}, \bibinfo {author} {\bibfnamefont {S.}~\bibnamefont {Zhang}}, \bibinfo
  {author} {\bibfnamefont {H.-F.}\ \bibnamefont {Song}}, \bibinfo {author}
  {\bibfnamefont {M.}~\bibnamefont {Troyer}}, \ and\ \bibinfo {author}
  {\bibfnamefont {A.~A.}\ \bibnamefont {Soluyanov}},\ }\bibfield  {title}
  {\enquote {\bibinfo {title} {{WannierTools}: An open-source software package
  for novel topological materials}}, }\href {\doibase
  10.1016/j.cpc.2017.09.033} {\bibfield  {journal} {\bibinfo  {journal}
  {Comput. Phys. Commun.}\ }\textbf {\bibinfo {volume} {224}},\ \bibinfo
  {pages} {405} (\bibinfo {year} {2018})}\BibitemShut {NoStop}%
\bibitem [{\citenamefont {Hao}\ \emph {et~al.}(2019)\citenamefont {Hao},
  \citenamefont {Liu}, \citenamefont {Feng}, \citenamefont {Ma}, \citenamefont
  {Schwier}, \citenamefont {Arita}, \citenamefont {Kumar}, \citenamefont {Hu},
  \citenamefont {Lu}, \citenamefont {Zeng}, \citenamefont {Wang}, \citenamefont
  {Hao}, \citenamefont {Sun}, \citenamefont {Zhang}, \citenamefont {Mei},
  \citenamefont {Ni}, \citenamefont {Wu}, \citenamefont {Shimada},
  \citenamefont {Chen}, \citenamefont {Liu},\ and\ \citenamefont
  {Liu}}]{prx.9.041038}%
  \BibitemOpen
  \bibfield  {author} {\bibinfo {author} {\bibfnamefont {Y.-J.}\ \bibnamefont
  {Hao}}, \bibinfo {author} {\bibfnamefont {P.}~\bibnamefont {Liu}}, \bibinfo
  {author} {\bibfnamefont {Y.}~\bibnamefont {Feng}}, \bibinfo {author}
  {\bibfnamefont {X.-M.}\ \bibnamefont {Ma}}, \bibinfo {author} {\bibfnamefont
  {E.~F.}\ \bibnamefont {Schwier}}, \bibinfo {author} {\bibfnamefont
  {M.}~\bibnamefont {Arita}},  \emph {et~al.},\ }\bibfield  {title} {\enquote
  {\bibinfo {title} {{Gapless surface Dirac cone in antiferromagnetic
  topological insulator ${\mathrm{MnBi}}_{2}{\mathrm{Te}}_{4}$}}}, }\href
  {\doibase 10.1103/PhysRevX.9.041038} {\bibfield  {journal} {\bibinfo
  {journal} {Phys. Rev. X}\ }\textbf {\bibinfo {volume} {9}},\ \bibinfo {pages}
  {041038} (\bibinfo {year} {2019})}\BibitemShut {NoStop}%
\bibitem [{\citenamefont {Xu}\ \emph {et~al.}(2019)\citenamefont {Xu},
  \citenamefont {Song}, \citenamefont {Wang}, \citenamefont {Weng},\ and\
  \citenamefont {Dai}}]{Xu2019PRL}%
  \BibitemOpen
  \bibfield  {author} {\bibinfo {author} {\bibfnamefont {Y.}~\bibnamefont
  {Xu}}, \bibinfo {author} {\bibfnamefont {Z.}~\bibnamefont {Song}}, \bibinfo
  {author} {\bibfnamefont {Z.}~\bibnamefont {Wang}}, \bibinfo {author}
  {\bibfnamefont {H.}~\bibnamefont {Weng}}, \ and\ \bibinfo {author}
  {\bibfnamefont {X.}~\bibnamefont {Dai}},\ }\bibfield  {title} {\enquote
  {\bibinfo {title} {{Higher-order topology of the axion insulator
  EuIn$_2$As$_2$}}}, }\href {\doibase 10.1103/physrevlett.122.256402}
  {\bibfield  {journal} {\bibinfo  {journal} {Phys. Rev. Lett.}\ }\textbf
  {\bibinfo {volume} {122}},\ \bibinfo {pages} {256402} (\bibinfo {year}
  {2019})}\BibitemShut {NoStop}%
\bibitem [{\citenamefont {Gui}\ \emph {et~al.}(2019)\citenamefont {Gui},
  \citenamefont {Pletikosic}, \citenamefont {Cao}, \citenamefont {Tien},
  \citenamefont {Xu}, \citenamefont {Zhong}, \citenamefont {Wang},
  \citenamefont {Chang}, \citenamefont {Jia}, \citenamefont {Valla},
  \citenamefont {Xie},\ and\ \citenamefont {Cava}}]{acscentsci.9b00202}%
  \BibitemOpen
  \bibfield  {author} {\bibinfo {author} {\bibfnamefont {X.}~\bibnamefont
  {Gui}}, \bibinfo {author} {\bibfnamefont {I.}~\bibnamefont {Pletikosic}},
  \bibinfo {author} {\bibfnamefont {H.}~\bibnamefont {Cao}}, \bibinfo {author}
  {\bibfnamefont {H.-J.}\ \bibnamefont {Tien}}, \bibinfo {author}
  {\bibfnamefont {X.}~\bibnamefont {Xu}}, \bibinfo {author} {\bibfnamefont
  {R.}~\bibnamefont {Zhong}},  \emph {et~al.},\ }\bibfield  {title} {\enquote
  {\bibinfo {title} {{A new magnetic topological quantum material candidate by
  design}}}, }\href {\doibase 10.1021/acscentsci.9b00202} {\bibfield  {journal}
  {\bibinfo  {journal} {ACS Cent. Sci.}\ }\textbf {\bibinfo {volume} {5}},\
  \bibinfo {pages} {900} (\bibinfo {year} {2019})}\BibitemShut {NoStop}%
\bibitem [{\citenamefont {Ma}\ \emph {et~al.}(2020)\citenamefont {Ma},
  \citenamefont {Wang}, \citenamefont {Nie}, \citenamefont {Yi}, \citenamefont
  {Xu}, \citenamefont {Li}, \citenamefont {Jandke}, \citenamefont {Wulfhekel},
  \citenamefont {Huang}, \citenamefont {West}, \citenamefont {Richard},
  \citenamefont {Chikina}, \citenamefont {Strocov}, \citenamefont {Mesot},
  \citenamefont {Weng}, \citenamefont {Zhang}, \citenamefont {Shi},
  \citenamefont {Qian}, \citenamefont {Shi},\ and\ \citenamefont
  {Ding}}]{adma.201907565}%
  \BibitemOpen
  \bibfield  {author} {\bibinfo {author} {\bibfnamefont {J.}~\bibnamefont
  {Ma}}, \bibinfo {author} {\bibfnamefont {H.}~\bibnamefont {Wang}}, \bibinfo
  {author} {\bibfnamefont {S.}~\bibnamefont {Nie}}, \bibinfo {author}
  {\bibfnamefont {C.}~\bibnamefont {Yi}}, \bibinfo {author} {\bibfnamefont
  {Y.}~\bibnamefont {Xu}}, \bibinfo {author} {\bibfnamefont {H.}~\bibnamefont
  {Li}},  \emph {et~al.},\ }\bibfield  {title} {\enquote {\bibinfo {title}
  {{Emergence of nontrivial low-energy Dirac fermions in antiferromagnetic
  EuCd$_2$As$_2$}}}, }\href {\doibase https://doi.org/10.1002/adma.201907565}
  {\bibfield  {journal} {\bibinfo  {journal} {Adv. Mater.}\ }\textbf {\bibinfo
  {volume} {32}},\ \bibinfo {pages} {1907565} (\bibinfo {year}
  {2020})}\BibitemShut {NoStop}%
\bibitem [{\citenamefont {Chen}\ \emph {et~al.}(2014)\citenamefont {Chen},
  \citenamefont {Niu},\ and\ \citenamefont
  {MacDonald}}]{PhysRevLett.112.017205}%
  \BibitemOpen
  \bibfield  {author} {\bibinfo {author} {\bibfnamefont {H.}~\bibnamefont
  {Chen}}, \bibinfo {author} {\bibfnamefont {Q.}~\bibnamefont {Niu}}, \ and\
  \bibinfo {author} {\bibfnamefont {A.~H.}\ \bibnamefont {MacDonald}},\
  }\bibfield  {title} {\enquote {\bibinfo {title} {{Anomalous Hall effect
  arising from noncollinear antiferromagnetism}}}, }\href {\doibase
  10.1103/PhysRevLett.112.017205} {\bibfield  {journal} {\bibinfo  {journal}
  {Phys. Rev. Lett.}\ }\textbf {\bibinfo {volume} {112}},\ \bibinfo {pages}
  {017205} (\bibinfo {year} {2014})}\BibitemShut {NoStop}%
\bibitem [{\citenamefont {Nakatsuji}\ \emph {et~al.}(2015)\citenamefont
  {Nakatsuji}, \citenamefont {Kiyohara},\ and\ \citenamefont
  {Higo}}]{Nakatsuji_2015}%
  \BibitemOpen
  \bibfield  {author} {\bibinfo {author} {\bibfnamefont {S.}~\bibnamefont
  {Nakatsuji}}, \bibinfo {author} {\bibfnamefont {N.}~\bibnamefont {Kiyohara}},
  \ and\ \bibinfo {author} {\bibfnamefont {T.}~\bibnamefont {Higo}},\
  }\bibfield  {title} {\enquote {\bibinfo {title} {{Large anomalous Hall effect
  in a non-collinear antiferromagnet at room temperature}}}, }\href {\doibase
  10.1038/nature15723} {\bibfield  {journal} {\bibinfo  {journal} {Nature}\
  }\textbf {\bibinfo {volume} {527}},\ \bibinfo {pages} {212} (\bibinfo {year}
  {2015})}\BibitemShut {NoStop}%
\end{thebibliography}%
\end{document}